\input harvmac
\input epsf
\newif\ifdraft\draftfalse
%\drafttrue
\newif\ifinter\interfalse
%\intertrue
\ifdraft\draftmode\else\interfalse\fi
\def\journal#1&#2(#3){\unskip, \sl #1\ \bf #2 \rm(19#3) }
\def\andjournal#1&#2(#3){\sl #1~\bf #2 \rm (19#3) }

\def\ie{{\it i.e.}}
\def\eg{{\it e.g.}}

\def\p{\partial}

\def\frac#1#2{{#1\over#2}}

\def\half{\frac12}

\def\inbar{\,\vrule height1.5ex width.4pt depth0pt}
\def\IC{\relax\hbox{$\inbar\kern-.3em{\rm C}$}}
\def\IR{\relax{\rm I\kern-.18em R}}
\def\IP{\relax{\rm I\kern-.18em P}}

%
%%%%%%%%%%%%%%%%%%%%%%%%%%%%%%%%%%%%
%

%\def\ap#1#2#3{Ann. Phys. {\bf #1} (#2) #3}

%
\catcode`\@=11
\def\slash#1{\mathord{\mathpalette\c@ncel{#1}}}
\overfullrule=0pt

\def\BB{{\cal B}}

\def\MM{{\cal M}}
\def\NN{{\cal N}}

\def\underrel#1\over#2{\mathrel{\mathop{\kern\z@#1}\limits_{#2}}}

\catcode`\@=12

%%%%%%%%%%%%%%%%%%%%%%%%%%%%%%%%%%%%%%%%%%%%%%%%%%%%%%%%%%%%%%

%

\def\tr{{\rm tr}}

%%%%%%%%%%%%%%%%%%%%%%%%%%%%%%%%%%%%%%%%%%%%%%%%%%%%%%%%%%%%%%
% new defs:

\def\[{[}
\def\]{]}

\def\comment#1{ }

%%%%%%%%%%%%%%%%%%%%%%%%%%%%%%%%%%%%%%%%%%%%%%%%%%%%%%%%%%%%%%
%%% Oskar's definitions:
%
%% A box for a short draft note
\def\draftnote#1{\ifdraft{\baselineskip2ex
                 \vbox{\kern1em\hrule\hbox{\vrule\kern1em\vbox{\kern1ex
                 \noindent \underbar{NOTE}: #1
             \vskip1ex}\kern1em\vrule}\hrule}}\fi}
%% A box for a short internal note
\def\internote#1{\ifinter{\baselineskip2ex
                 \vbox{\kern1em\hrule\hbox{\vrule\kern1em\vbox{\kern1ex
                 \noindent \underbar{Internal Note}: #1
             \vskip1ex}\kern1em\vrule}\hrule}}\fi}
%% A few internal words

%
%% Greek letters
%
%\def\al{\alpha}
%\def\bt{\beta}
%\def\gm{\gamma}                \def\Gm{\Gamma}
%\def\dl{\delta}                \def\Dl{\Delta}
%\def\ep{\epsilon}
%\def\vep{\varepsilon}

%\def\io{\iota}
%\def\kp{\kappa}
%\def\lm{\lambda}               \def\Lm{\Lambda}
%%\mu,\nu unchanged

%\def\th{\theta}               \def\Th{\Theta}
%\def\vth{\vartheta}
%%\phi unchanged               \Phi unchanged
%\def\vph{\varphi}
%%\psi unchanged               \Psi unchanged
%%\chi unchanged

%\def\om{\omega}               \def\Om{\Omega}
%%\pi unchanged                \Pi unchanged
%\def\vpi{\varpi}
%%\rho unchanged
%\def\vro{\varrho}
%\def\sg{\sigma}               \def\Sg{\Sigma}
%\def\vsg{\varsigma}
%%\tau unchanged
%\def\up{\upsilon}             \Up{\Upsilon}
%%\xi unchanged                \Xi unchanged
%%\eta unchanged
%\def\zt{\zeta}
%
%% Capital roman double letters (blackboard font)
%
%\def\inbar{\hskip.3em\vrule height1.5ex width.4pt depth0pt}
%\def\IC{\relax{\inbar\kern-.3em{\rm C}}}
%\def\IN{\relax{\rm I\kern-.16em N}}
%\def\IP{\relax{\rm I\kern-.18em P}}
%\def\IQ{\relax\hbox{$\inbar$\kern-.3em{\rm Q}}}
%\def\IR{\relax{\rm I\kern-.18em R}}
%\def\IZ{\relax{\rm Z\kern-.8em Z}}
%
%% Other Defs
%

%

%mydef

\def\p{\partial}
\def\al{\tilde a}
\def\ale{{\tilde a^{el}}}
\def\am{{\tilde a}^{m}}
\def\x{{\tilde x}}
\def\tr{{\rm tr}}
\def\Tr{{\rm Tr}}

\def\y{{\tilde y}}

%\IntriligatorJJ
\lref\IntriligatorJJ{
K.~Intriligator and B.~Wecht,
``The exact superconformal R-symmetry maximizes a,''
arXiv:hep-th/0304128.
%%CITATION = HEP-TH 0304128;%%
}

%\CardyCW
\lref\CardyCW{
J.~L.~Cardy,
``Is There A C Theorem In Four-Dimensions?,''
Phys.\ Lett.\ B {\bf 215}, 749 (1988).
%%CITATION = PHLTA,B215,749;%%
}
%\JackEB
\lref\JackEB{
I.~Jack and H.~Osborn,
``Analogs For The C Theorem For Four-Dimensional Renormalizable Field Theories,''
Nucl.\ Phys.\ B {\bf 343}, 647 (1990).
%%CITATION = NUPHA,B343,647;%%
}
%\CappelliYC
\lref\CappelliYC{
A.~Cappelli, D.~Friedan and J.~I.~Latorre,
``C Theorem And Spectral Representation,''
Nucl.\ Phys.\ B {\bf 352}, 616 (1991).
%%CITATION = NUPHA,B352,616;%%
}
%\CappelliKE
\lref\CappelliKE{
A.~Cappelli, J.~I.~Latorre and X.~Vilasis-Cardona,
``Renormalization group patterns and C theorem in more than two-dimensions,''
Nucl.\ Phys.\ B {\bf 376}, 510 (1992)
[arXiv:hep-th/9109041].
%%CITATION = HEP-TH 9109041;%%
}
%\BastianelliVV
\lref\BastianelliVV{
F.~Bastianelli,
``Tests for C-theorems in 4D,''
Phys.\ Lett.\ B {\bf 369}, 249 (1996)
[arXiv:hep-th/9511065].
%%CITATION = HEP-TH 9511065;%%
}

%\ForteDX
\lref\ForteDX{
S.~Forte and J.~I.~Latorre,
``A proof of the irreversibility of renormalization group flows in four  dimensions,''
Nucl.\ Phys.\ B {\bf 535}, 709 (1998)
[arXiv:hep-th/9805015].
%%CITATION = HEP-TH 9805015;%%
}

%\AnselmiYS
\lref\AnselmiYS{
D.~Anselmi, J.~Erlich, D.~Z.~Freedman and A.~A.~Johansen,
``Positivity constraints on anomalies in supersymmetric gauge theories,''
Phys.\ Rev.\ D {\bf 57}, 7570 (1998)
[arXiv:hep-th/9711035].
%%CITATION = HEP-TH 9711035;%%
}

%\AnselmiAM
\lref\AnselmiAM{
D.~Anselmi, D.~Z.~Freedman, M.~T.~Grisaru and A.~A.~Johansen,
``Nonperturbative formulas for central functions of supersymmetric gauge  theories,''
Nucl.\ Phys.\ B {\bf 526}, 543 (1998)
[arXiv:hep-th/9708042].
%%CITATION = HEP-TH 9708042;%%
}

%\AnselmiUK
\lref\AnselmiUK{
D.~Anselmi,
``Quantum irreversibility in arbitrary dimension,''
Nucl.\ Phys.\ B {\bf 567}, 331 (2000)
[arXiv:hep-th/9905005].
%%CITATION = HEP-TH 9905005;%%
}
%\CappelliDV
\lref\CappelliDV{
A.~Cappelli, G.~D'Appollonio, R.~Guida and N.~Magnoli,
``On the c-theorem in more than two dimensions,''
arXiv:hep-th/0009119.
%%CITATION = HEP-TH 0009119;%%
}

%\NovikovUC
\lref\NovikovUC{
V.~A.~Novikov, M.~A.~Shifman, A.~I.~Vainshtein and V.~I.~Zakharov,
``Exact Gell-Mann-Low Function Of Supersymmetric Yang-Mills Theories From Instanton Calculus,''
Nucl.\ Phys.\ B {\bf 229}, 381 (1983).
%%CITATION = NUPHA,B229,381;%%
}
%\ShifmanZI
\lref\ShifmanZI{
M.~A.~Shifman and A.~I.~Vainshtein,
``Solution Of The Anomaly Puzzle In Susy Gauge Theories And The Wilson Operator Expansion,''
Nucl.\ Phys.\ B {\bf 277}, 456 (1986)
[Sov.\ Phys.\ JETP {\bf 64}, 428 (1986\ ZETFA,91,723-744.1986)].
%%CITATION = NUPHA,B277,456;%%
}

%\MackJE
\lref\MackJE{
G.~Mack,
``All Unitary Ray Representations Of The Conformal Group SU(2,2) With Positive Energy,''
Commun.\ Math.\ Phys.\  {\bf 55}, 1 (1977).
%%CITATION = CMPHA,55,1;%%
}

%\SeibergPQ
\lref\SeibergPQ{
N.~Seiberg,
``Electric - magnetic duality in supersymmetric nonAbelian gauge theories,''
Nucl.\ Phys.\ B {\bf 435}, 129 (1995)
[arXiv:hep-th/9411149].
%%CITATION = HEP-TH 9411149;%%
}

%\KutasovVE
\lref\KutasovVE{
D.~Kutasov,
``A Comment on duality in N=1 supersymmetric nonAbelian gauge theories,''
Phys.\ Lett.\ B {\bf 351}, 230 (1995)
[arXiv:hep-th/9503086].
%%CITATION = HEP-TH 9503086;%%
}
%\KutasovNP
\lref\KutasovNP{
D.~Kutasov and A.~Schwimmer,
``On duality in supersymmetric Yang-Mills theory,''
Phys.\ Lett.\ B {\bf 354}, 315 (1995)
[arXiv:hep-th/9505004].
%%CITATION = HEP-TH 9505004;%%
}
%\KutasovSS
\lref\KutasovSS{
D.~Kutasov, A.~Schwimmer and N.~Seiberg,
%``Chiral Rings, Singularity Theory and Electric-Magnetic Duality,''
Nucl.\ Phys.\ B {\bf 459}, 455 (1996)
[arXiv:hep-th/9510222].
%%CITATION = HEP-TH 9510222;%%
}
%\AffleckMK
\lref\AffleckMK{
I.~Affleck, M.~Dine and N.~Seiberg,
``Dynamical Supersymmetry Breaking In Supersymmetric QCD,''
Nucl.\ Phys.\ B {\bf 241}, 493 (1984).
%%CITATION = NUPHA,B241,493;%%
}
%\AffleckXZ
\lref\AffleckXZ{
I.~Affleck, M.~Dine and N.~Seiberg,
``Dynamical Supersymmetry Breaking In Four-Dimensions And Its Phenomenological Implications,''
Nucl.\ Phys.\ B {\bf 256}, 557 (1985).
%%CITATION = NUPHA,B256,557;%%
}
%\ZamolodchikovGT
\lref\ZamolodchikovGT{
A.~B.~Zamolodchikov,
%``'Irreversibility' Of The Flux Of The Renormalization Group In A 2-D Field Theory,''
JETP Lett.\  {\bf 43}, 730 (1986)
[Pisma Zh.\ Eksp.\ Teor.\ Fiz.\  {\bf 43}, 565 (1986)].
%%CITATION = JTPLA,43,730;%%
}

%%%%%%%%%%%%%%%%%%%%%%%%%%%%%%%%%%%%%%%%%%%%%%%%%%%
{\Title{\vbox{
%\hbox{hep--th/}
\hbox{EFI-03-40}}}
{\vbox{\centerline{Central Charges and  $U(1)_R$ Symmetries in}
\vskip 10pt
\centerline{${\cal N}=1$ Super Yang-Mills}}}

\bigskip
\centerline{David Kutasov,
Andrei Parnachev and David A. Sahakyan}
\bigskip
\centerline{\it Department of Physics, University of Chicago}
\centerline{\it 5640 S. Ellis Av., Chicago, IL 60637, USA}
\centerline{kutasov, andrei, sahakian@theory.uchicago.edu}

\vskip 0.5 cm

We use recent results of Intriligator and Wecht \IntriligatorJJ\
to study the phase structure of $\NN=1$ super Yang-Mills theory
with gauge group $SU(N_c)$, a chiral superfield in the adjoint,
and $N_f$ chiral superfields in the fundamental representation 
of the gauge group. Our discussion sheds new light on \IntriligatorJJ\ and
supports the conjecture that the central charge $a$ decreases under
RG flows and is non-negative in unitary four dimensional conformal
field theories.

\Date{August 2003}

%\nopagenumber
\vfill
\eject}

\newsec{Introduction}

Recently, K. Intriligator and B. Wecht \IntriligatorJJ\ proposed a
solution to an old problem, of determining the $U(1)_R$ charges
of chiral operators at non-trivial fixed points of the Renormalization
Group (RG) in $\NN=1$ supersymmetric four dimensional gauge
theories. The results of \IntriligatorJJ\ also provide support 
for the conjectured ``$a$-theorem''
\refs{\CardyCW\JackEB\CappelliYC\CappelliKE\BastianelliVV\AnselmiAM\AnselmiYS\ForteDX\AnselmiUK
-\CappelliDV}, which states\foot{The $a$-theorem
actually refers to an anomaly associated with the stress-tensor,
but in $\NN=1$ superconformal theories it is related by supersymmetry
to the anomaly associated with the R-current discussed here.}
that the combination of `t Hooft anomalies
\eqn\adef{a\equiv {3\over 32} (3\tr R^3-\tr R)\equiv {3\over 32}\tilde a~,}
is always positive and lower at an IR fixed point of an RG flow than at the corresponding
UV fixed point. Here $R$ is the $U(1)_R$ charge which belongs to the
$\NN=1$ superconformal multiplet at a fixed point of the RG, and the trace
runs over the chiral fermions in the multiplets. We will mostly refer below
to the quantity $\tilde a$ defined in \adef, which differs from $a$ by the factor
$3/32$.

As we review below, the analysis of \IntriligatorJJ\ leaves some open questions that
need to be studied on a case by case basis, such as the range of validity of the
results, and the implications for the $a$-theorem. In this note we apply the
results of \IntriligatorJJ\ to $\NN=1$ super Yang-Mills (SYM) coupled to
a single chiral superfield $X$ in the adjoint representation of the gauge group
$SU(N_c)$, and $N_f$ ``flavors'' of chiral superfields in the fundamental
representation of the gauge group, $Q^i$, $\tilde Q_{\tilde i}$, $i,\tilde i=1
\cdots N_f$. This class of theories exhibits a rich pattern of RG flows, some of
which are understood, but there are important open questions. In particular, it
was pointed out in \AnselmiYS\ that some of the known RG flows might lead to
counter-examples to the $a$-theorem, depending on some detailed features
of the flows that were not understood at the time.

Thus, this class of theories is a good testing ground for the techniques of Intriligator
and Wecht (IW). We will see that the results of \IntriligatorJJ\ allow one to obtain a more detailed
picture of the phase structure. In the process, we will get new insights into the construction
of \IntriligatorJJ, and the validity of the $a$-theorem. To set the stage, we start with a general
discussion of RG flows and fixed points in $\NN=1$ SYM.

An asymptotically free gauge theory describes an RG flow between a free theory
in the UV, where the gauge coupling vanishes, and an interacting theory in the
IR, where the gauge coupling is non-zero. Moreover, it might happen that an operator
that is irrelevant (in the RG sense) near the free UV fixed point becomes marginal
or relevant near the IR fixed point, and leads to new deformations that allow one to flow
further in the space of couplings and explore additional fixed points. The program of
``solving'' a gauge theory involves understanding all the fixed points that can be reached
from the free UV theory  this way, and then understanding the theory along the RG flows
that connect the different fixed points.

At a fixed point of the RG, the theory becomes scale invariant, and in all known examples
conformal. We will be interested in $\NN=1$ supersymmetric theories, which become
superconformal at fixed points. The $\NN=1$ superconformal algebra contains a $U(1)_R$
charge $R$, which determines the scaling dimensions of chiral operators, via the relation
\eqn\delt{
\Delta({\cal O})={3\over 2}R({\cal O})~.}
Thus, determining the $R$-charges of chiral operators at fixed points of the RG is important,
since it leads to a determination of their scaling dimensions at these fixed points.

One general idea that is known to be useful for identifying the $U(1)_R$ charge $R$ is to
postulate that the current $J_\mu$ that becomes part of the superconformal multiplet in the
IR is conserved throughout the RG flow, and thus can be identified already in the vicinity of
the (asymptotically free) UV fixed point. Strong support for this idea is provided by
the form of the NSVZ $\beta$ function \refs{\NovikovUC, \ShifmanZI} of $\NN=1$ SYM with
gauge group $G$ and chiral superfields $\Phi_i$ in the representations $r_i$ of the gauge group:
\eqn\nsvzbeta{\beta(\alpha)=
-{\alpha^2\over2\pi}{3T(G)-\sum_i T(r_i)(1-\gamma_i)\over 1-{\alpha\over2\pi}T(G)}~.}
Here $\alpha=g^2/4\pi$, $g$ is the gauge coupling, $\gamma_i$ is the anomalous dimension
of $\Phi_i$, $T(r)$ is the quadratic Casimir corresponding to the representation $r$,
${\rm Tr}_r(T^aT^b)=T(r)\delta^{ab}$, and $T(G)=T(r={\rm adjoint})$.
If the coupling $\alpha$ in the IR is sufficiently small that we do not have to worry about
the denominator in \nsvzbeta, non-trivial fixed points of the RG correspond to a zero of
the numerator,
\eqn\nontriv{3T(G)-\sum_i T(r_i)(1-\gamma_i)=0~.}
At a fixed point, the scaling dimension of  $\Phi_i$ is given by
$\Delta_i=1+\half\gamma_i$. Using the relation between the dimension and the
R-charge, \delt, one can rewrite the condition for a fixed point \nontriv\
as
\eqn\anomfree{T(G)+\sum_i T(r_i)(R_i-1)=0~.}
This equation is also the condition that the R-symmetry
with  $R(\Phi_i)=R_i$  be anomaly free. Thus, it is natural to
postulate that the R-symmetry of the IR fixed point is one of the
anomaly free R-symmetries which satisfy \anomfree, and are therefore
conserved in the full theory. Indeed, the trace of the stress-tensor $T_\mu^\mu$
and the divergence of the $U(1)_R$ current $\partial_\mu J^\mu$ are in the same
supersymmetry multiplet. Thus, one expects the condition that $\beta(\alpha)=0$
(or $T_\mu^\mu=0$) to be related by supersymmetry to the condition that
the $U(1)_R$ current is anomaly free and hence conserved.

On general grounds, one does not expect that the technique described above
should be always valid. It might be that the $U(1)_R$ that becomes part of the
superconformal algebra in the extreme IR is an accidental symmetry of the IR
theory and is not visible from the UV. The argument above suggests that
this does not happen when the theory is sufficiently weakly coupled, and only
becomes an issue when the IR fixed point is ``too far'' from the UV fixed point.
{}From the point of view of \nsvzbeta, one way this might happen is if $\alpha$
exceeds in the IR the value for which the $\beta$ function has a pole.

There are actually two different ways in which such a violation might manifest
itself:

\item{(1)} Suppose that we found a candidate R-symmetry by solving \anomfree, and
it predicts that a particular gauge invariant chiral operator $M$ has R-charge
$R(M)<2/3$. This is inconsistent with unitarity \MackJE. In that case, it is believed
that what happens is the following \SeibergPQ. The correct answer is $R(M)=2/3$, $M$
is a free field in the IR CFT, and the correct R-symmetry is a combination
of the solution of \anomfree\ and an accidental symmetry of the IR theory
which acts only on the free field $M$.

\item{(2)} Even if the candidate R-symmetry assigns R-charge larger than 2/3
to gauge invariant chiral superfields, it may be invalid. This is in a sense
more  problematic, since unlike case (1), there is no obvious ``smoking gun'',
and no general procedure for fixing the problem.

\noindent
To illustrate the above general considerations, consider the case of supersymmetric
QCD, with gauge group $SU(N_c)$ and $N_f$ fundamental multiplets, $Q^i$,
$\tilde Q_{\tilde i}$.
The theory is asymptotically free for $N_f<3N_c$. Following the logic outlined above,
one looks for the R-symmetry in the IR among the anomaly free symmetries of the full
theory. Since the dynamics is invariant under interchange of $Q$ and $\tilde Q$, one must
have $R(Q)=R(\tilde Q)$. There is then a unique solution to the anomaly 
constraint\foot{For $SU(N_c)$, $T({\rm fund})=1/2$, $T({\rm adjoint})=N_c$.}
\anomfree:
\eqn\quarkIR{
R(Q)=R(\tilde Q)={N_f-N_c\over N_f}~.}
The gauge invariant chiral operators
\eqn\ginv{
\eqalign{
{\cal M}^{i}_{\tilde i}=&Q^i\tilde Q_{\tilde i}~,\cr
{\cal B}^{[i_1,\cdots,i_{N_c}]}=&Q^{i_1}\cdots Q^{i_{N_c}}~,\cr
{\tilde {\cal B}}_{[\tilde i_1,\cdots,\tilde i_{N_c}]}=&\tilde Q_{\tilde i_1}\cdots \tilde Q_{\tilde i_{N_c}}~,
}}
are assigned R-charges
\eqn\ginvir{\eqalign{
&R({\cal M})=2{(N_f-N_c)\over N_f}~,\cr
&R({\cal B})=R(\tilde{ \cal B})={N_c(N_f-N_c)\over N_f}~.
}}
Eq. \ginvir\ is known to break down at $N_f=3N_c/2$, illustrating
both of the phenomena mentioned above. First, for $N_f<3N_c/2$,
$R(\MM)<2/3$, and as explained above  the chiral
superfield $\MM$ \ginv\ becomes free in the IR. This is an example
of point (1) above.

One might expect that the prediction for $R(\BB)$ in \ginvir\ should still
be valid, since it is typically large in the range of $N_f,N_c$ under
consideration, but this is known to be incorrect. In fact, the
right description for $N_f<3N_c/2$ is in terms of a Seiberg dual 
theory \SeibergPQ, with gauge group $SU(N_f-N_c)$. The baryons $\BB$ 
\ginv\ can be expressed in terms of the magnetic quarks $q$ as 
$\BB\sim q^{N_f-N_c}$. Since the latter are free for 
$N_c<N_f \le 3N_c/2$, one has
\eqn\baryonout
{R({\cal B})=2(N_f-N_c)/3~.}
We see that \ginvir\ fails for $N_f<3N_c/2$, even though $R(\BB)$ is typically
large and positive when that happens. This is an example of point (2) above.
It is important to emphasize that while in this case both of the kinds
of violations discussed in points (1) and (2) above occur in the same
regime, $N_f<3N_c/2$, in general these two types of phenomena are distinct,
and we will see examples later where only one or the other occurs.

To summarize, we see that there is a finite region, $3N_c/2\le N_f\le 3N_c$
in which one can identify the IR R-charge as a symmetry of the full theory.
For $N_f<3N_c/2$ this idea fails, but then Seiberg duality comes to the rescue
and allows one to solve the problem using a weakly coupled description.

In more general SYM theories the situation is expected to be qualitatively similar
to that described above. Imagine, for simplicity, that there is a parameter, like
$N_f$, as a function of which the IR coupling varies between zero and some
finite (or infinite) value. Then, one expects to find a range of parameters for which
the IR $U(1)_R$ is one of the anomaly free symmetries \anomfree, perhaps corrected
by taking into account the decoupling of some free fields, as in point (1) above.
Beyond this range, this method breaks down and one needs to proceed
in some other way (\eg\ use Seiberg duality).

The first step of this process, finding the solution of \anomfree\ that corresponds
to the IR $U(1)_R$, is in general non-trivial since the solution is not unique.
For example, in the case of interest in this paper, adjoint SQCD, \ie\
SYM with gauge group $G=SU(N_c)$
and matter superfields $X$, $Q^i$, $\tilde Q_{\tilde i}$ in the adjoint, fundamental and
anti-fundamental representations of $G$, respectively $(i,\tilde i=1,\cdots, N_f)$,
assigning R-charge $R(Q)$ to $Q$, $\tilde Q$ and $R(X)$ to $X$, one has \anomfree:
\eqn\nfcrq{N_fR(Q)+N_cR(X)=N_f~.}
Thus, there is a one parameter set of candidate R-symmetries, and it is not clear
which of these is the correct one. This is the problem solved by IW \IntriligatorJJ.
These authors proved that if the IR R-charge is a solution of \anomfree, it is the one
that locally maximizes $a$ \adef\ over the set of all solutions of \anomfree.~\foot{As
we will see below, the situation is actually slightly more involved.}

{}From the point of view of the general discussion above, a few natural questions are:
\item{(a)} Does the R-symmetry found by IW describe
the IR fixed point for all $N_f$, or does it fail 
for $N_f<N_f^*$, with some $N_f^*>0$?

\item{(b)} What does the R-symmetry of IW
imply for the possible flows corresponding to
relevant deformations of the IR fixed point?

\item{(c)} Are the RG flows implied by the IW results consistent
with the $a$-theorem? 

\noindent
The purpose of this paper is to discuss these issues. We will
find that the results of \IntriligatorJJ\ (corrected slightly to take into account
unitarity constraints) lead to a sensible picture of the structure
of RG flows in adjoint SQCD, which is in particular consistent 
with the results of \refs{\KutasovVE\KutasovNP-\KutasovSS} on 
Seiberg duality in these models, and with the $a$-theorem. 
The potential violations of the $a$-theorem pointed out in 
\AnselmiYS, as well as others, are avoided by these flows.
In the theory with vanishing superpotential for $X$, $Q^i$,
$\tilde Q_{\tilde i}$ we do not find any evidence for the breakdown
of the results of \IntriligatorJJ\ for any $N_f>0$, while in the theories
with a non-zero superpotential we exhibit examples in
which such an analysis has a limited domain of validity,
and in order to explore the whole phase diagram one has to 
appeal to Seiberg duality. 

The plan of the paper is as follows. In section 2 we study adjoint SQCD
with vanishing superpotential for large $N_f, N_c$, with fixed $N_c/N_f$. 
We apply the analysis of IW to this case, taking into account unitarity
constraints which modify slightly the results given in \IntriligatorJJ,
compute the R-charges of $Q$ and $X$, and the resulting central charge 
$\tilde a$. We show that the $a$-theorem is satisfied for the RG flows
in this system. 

In section 3 we study deformations of adjoint SQCD corresponding to
Higgsing the gauge group and turning on a polynomial superpotential
for the adjoint superfield $X$. Again, the $a$-theorem appears to be
satisfied, rather non-trivially.

In section 4 we discuss the dynamics of the system in the presence
of a polynomial superpotential, by using a dual description due to
\refs{\KutasovVE-\KutasovSS}. We show that the results of section
3 lead to a picture consistent with the duality. At the same time,
the duality predicts that the calculation of the central charge $\tilde a$
done in section 3 breaks down at some critical value of $N_c/N_f$, which
we find. Beyond that point one must switch to the dual variables in order to
compute it correctly.

Section 5 contains a brief discussion. In appendix A we derive some
technical results that are used in the text.

\newsec{Adjoint SQCD with vanishing superpotential}

In this section we will study adjoint SQCD, which was mentioned
in the introduction, using the results of \IntriligatorJJ.  For simplicity,
we will work in the large $N$ limit\foot{It is not difficult to extend the
analysis to finite $N_c$, $N_f$.}
\eqn\limit{
N_c>>1;\quad N_f>>1;\quad x\equiv {N_c\over N_f}\,\, =\,\, {\rm fixed}~,
}
and study the phase structure as a function of the continuous parameter $x$. The theory
is asymptotically free for $x>1/2$, and we will mostly restrict our discussion to this regime.

Under RG flow, the gauge theory in question flows in the IR to a non-trivial fixed point.
To find the $U(1)_R$ symmetry at that fixed point we follow \IntriligatorJJ.
We assign R-charge $y$ to $Q$, $\tilde Q$, and compute the ``trial'' central charge
\adef, which we will denote by $\tilde a^{(0)}(x,y)$, 
using \nfcrq\ to express the R-charge of $X$
as a function of $y$, $R(X)=(1-y)/x$. One finds:
\eqn\naiv{
\tilde a^{(0)}(x,y)/N_f^2=6x(y-1)^3-2x(y-1)+3x^2\left({1-y\over x}-1\right)^3-
x^2\left({1-y\over x}-1\right)+2x^2~.}
As shown in \IntriligatorJJ, the IR $U(1)_R$ can be determined by requiring
that the trial central charge $\tilde a^{(0)}(x,y)$ \naiv\ is at a local maximum
with respect to $y$. This leads to:
\eqn\naivR{
\eqalign{
R^{(0)}(Q)=&y^{(0)}={3+x(-3-6x+\sqrt{20x^2-1})\over 3-6x^2}~,\cr
%&\cr
R^{(0)}(X)=&{1-y^{(0)}\over x}={10\over 3(3+\sqrt{20x^2-1})}~.\cr}
}
Plugging \naivR\ into \naiv, one finds the following expression for the
central charge:
\eqn\aattii{\tilde a^{(0)}(x)=
{2N_f^2x^2\over 9(1-2x^2)^2}\left[18-90x^2+(20x^2-1)^{3/2}\right]~.}
Actually, \naivR, \aattii\ are not the full story, since one needs to take into account
unitarity constraints. Consider, for example, the chiral superfield ${\cal M}_1=\tilde QQ$.
According to \naivR, it violates the unitarity bound
for $x>3+\sqrt{7}$, and the procedure of \IntriligatorJJ\ must be modified in this regime.

As discussed in the introduction, it is believed that in this situation the infrared SCFT
splits into an (in general) interacting theory and a decoupled free superfield $\MM_1$. The trial
central charge $\tilde a^{(0)}$ \naiv\ can then be written as a sum of two contributions. One comes
from the decoupled superfield $\MM_1$,
\eqn\atildem{
\tilde a_M(R(\MM_1))/N_f^2=3\left[R(\MM_1)-1\right]^3-\left[R(\MM_1)-1\right]~,
}
where $R(\MM_1)=2R(Q)=2y$; the other contribution is due to the interacting SCFT,
\eqn\ainteract{
\tilde a_{\rm interact}=\tilde a^{(0)}(x,y)-\tilde a_M(R(\MM_1))~.
}
It is clear that in order to find the IR $U(1)_R$ we only have to extremize
$\tilde a_{\rm interact}$, since we know what to do with the free superfield $\MM_1$.
Thus, for $x>3+\sqrt{7}$, the results \naivR\ are invalid, and are replaced by those
following from the extremization of \ainteract. The full central charge is \AnselmiYS
\eqn\atotal{
\tilde a^{(1)}=\tilde a_{\rm interact}+\tilde a_M\left({2\over3}\right)~.
}
This is not the end of the story either, since at some yet larger value of
$x$, the operator $\MM_2=\tilde Q XQ$ reaches R-charge $2/3$, and the same
procedure has to be repeated for it. More generally, every time the R-charge of a gauge 
invariant chiral
operator in the theory drops below 2/3, the procedure of \IntriligatorJJ\ has to
be modified accordingly.

In practice one can proceed as follows. Denote by $M$ the collection of gauge invariant
chiral superfields whose R-charge is  smaller than $2/3$ for a particular value of $x$ (or
$N_f$). The trial central charge that one should extremize is
\eqn\ansel{\eqalign{
\al= &\al^{(0)}(x,y)+\sum_M\left[\al_M\left ({2\over 3}\right)-\al_M(R(M))\right]\cr
=&\al^{(0)}(x,y)+{1\over 9}\sum_M {\rm dim}(M)\left[2-3R(M)\right]^2\left[5-3R(M)\right]~,
}}
where $R(M)$ is the R-charge of $M$ under the trial R-symmetry, dim$(M)$ is the number
of fields with the same R-charge, and the sum over $M$ runs over fields with different
R-charges. It is easy to generalize the above discussion to other gauge groups 
and matter contents.

In adjoint SQCD there are three types of gauge invariant chiral superfields that are relevant
for the preceding discussion:
\eqn\ginv{
\eqalign{
&{\cal B}^{(n_1,n_2,\cdots,n_k)}=Q^{n_1}_{(1)}\cdots Q^{n_k}_{(k)};\qquad 
\sum_{l=1}^kn_l=N_c;\quad n_l\leq N_f;\quad
k=1,2,\cdots~,\cr
&\tr X^j,\;\;j=2,3,\cdots~,\cr
&({\cal M}_j)^i_{\tilde i}=\tilde Q_{\tilde i} X^{j-1} Q^i,\quad j=1,2,\cdots~,
}
}
where in the first line the color indices are contracted with an $\epsilon$ tensor,
and following \KutasovNP\ we introduced ``dressed quarks''
\eqn\qx{
Q_{(l)}=X^{l-1}Q;\quad l=1,2,\cdots~.
}
There are also baryons $\tilde \BB$ obtained from \ginv\  by replacing $Q\to\tilde Q$.
The R-charge of the baryons $\cal B$ is given by
\eqn\rchrgb{
R({\cal B}^{(n_1,n_2,\cdots,n_k)})=\sum_{l=1}^k n_l(l-1)R(X)+N_c R(Q)~.
}
In the limit we are considering \limit, the R-charge of baryons is positive and infinite
since,  as we will see later, $R(X)$ and $R(Q)$ are both positive. 
Therefore, $\BB$ and $\tilde \BB$ do not
contribute to the correction terms in \ansel. The fields on the second line of \ginv\ do not
contribute to these corrections either, even if their R-charge reaches $2/3$, since in the limit
\limit\ their contribution to $\al$ is smaller than the other terms  by a factor of $N_f^2$.
On the other hand the fields $({\cal M}_j)_{\tilde i}^i$  can potentially change $\al$ in the
limit that we are considering, since there are $N_f^2$ of them for each $j$. The R-charge
of $({\cal M}_j)^i_{\tilde i}$ computed as a composite field is given by
\eqn\mes{
R({\cal M}_j)=2y+(j-1){1-y\over x}~.
}
We will see later that for large $x$, $y$ approaches a constant $y_0<1/3$. Thus,
for any given $j$ there exists a value of $x$ above which $\MM_j$ becomes free.

In order to determine $\al$, it is convenient to introduce an auxiliary quantity,
the trial central charge computed with the assumption that the first $p(=0,1,2,\cdots)$ 
meson fields
$\MM_1,\cdots,\MM_p$ are free:
\eqn\correct{
\eqalign{
&\al^{(p)}(x,y)/N_f^2
=6x(y-1)^3-2x(y-1)+3x^2\left({1-y\over x}-1\right)^3-
x^2\left({1-y\over x}-1\right)+2x^2+\cr
&\cr
&{1\over 9}\sum_{j=1}^p
\left[2- 3\left(2y + (j - 1){1 - y\over x}\right)\right]^2
\left[5 - 3\left(2y + (j - 1){1 - y\over x}\right)\right]~.
}
}
The central charge $\al$ can then be determined via the following process:
\item{(1)} Maximize each $\al^{(p)}$ with respect to $y$, 
and find the corresponding value of $y$, $y^{(p)}(x)$.
\item{(2)} Substitute $y^{(p)}$ into \mes\ and compute $\tilde p(p,x)$, such that
$\MM_1\cdots\MM_{\tilde p}$ have $R\leq 2/3$, but $R(\MM_{\tilde p+1})>2/3$.
\item{(3)} There is a unique value of $p$, $p_0$, for which\foot{Note that $p_0$
depends on $x$.} $\tilde p(p_0,x)=p_0$.
The central charge $\al(x)$ is given by $\al(x)=\al^{(p_0)}(x,y^{(p_0)}(x))$.

\noindent
It is useful to note that according to \correct, 
the $p$-th meson $\MM_p$ becomes free when
\eqn\alep{
\eqalign{
&\al^{(p-1)}=\al^{(p)},\cr
&\p_y \al^{(p-1)}=\p_y \al^{(p)}~.
}
}
Therefore, the R-charges computed from $\al^{(p-1)}$, $\al^{(p)}$ coincide at this
point. Moreover, at this point
\eqn\alepo{\eqalign{
&\al^{(p-1)}(x,y^{(p-1)}(x))=\al^{(p)}(x,y^{(p)}(x))~,\cr
&{d\over dx} \al^{(p-1)}(x,y^{(p-1)}(x))={d\over dx}\al^{(p)}(x,y^{(p)}(x))~,
}
}
which means that the central charge $\al$ is a continuous, smooth function of $x$. 
One can set up an iterative algorithm for finding the central charge
$\al(x)$ and the R-charges $R(Q)$ and $R(X)$ as follows.  
Start with $\al^{(0)}$ and find the
value of $x$ for which the R-charge of $\MM_1=\tilde Q Q$ approaches $2/3$. At that
point  ${\cal M}_1$ becomes free and we have to switch to the $\al^{(1)}$ description. 
Then look for the value of $x$ at which ${\cal M}_2$ becomes free
and decouples, switch to $\al^{(2)}$, etc. This process can be obviously 
continued to arbitrarily large $x$.

\midinsert\bigskip{\vbox{{\epsfxsize=3in
        \nobreak
    \centerline{\epsfbox{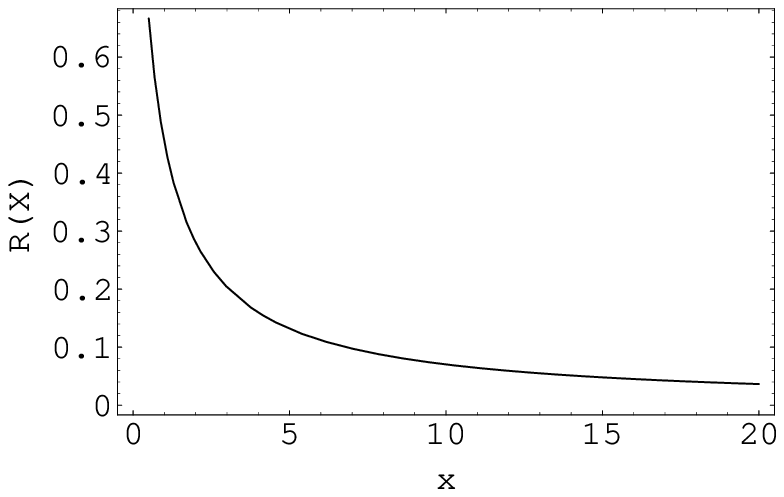}}
        \nobreak\bigskip
    {\raggedright\it \vbox{
{\bf Fig 1.}
{\it The R-charge of $X$ as a function of $x$.}}}}}}
\bigskip\endinsert

\midinsert\bigskip{\vbox{{\epsfxsize=3in
        \nobreak
    \centerline{\epsfbox{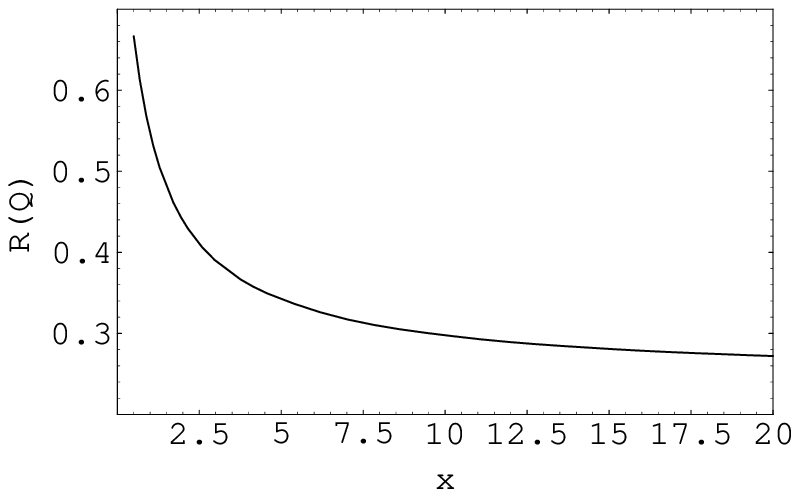}}
        \nobreak\bigskip
    {\raggedright\it \vbox{
{\bf Fig 2.}
{\it The R-charge of $Q$ as a function of $x$. }}}}}}
\bigskip\endinsert

We have implemented this algorithm using Mathematica. The results for the R-charges
are given in figures 1, 2. A few comments are in order regarding this procedure:
\item{(1)} It assumes that the mesons become free sequentially,
\ie\ at the point when $\MM_p$ becomes free, all the mesons
$\MM_j$ with $j<p$ are already free. This is indeed the case 
provided $R(X)>0$, as one can see from \mes.
\item{(2)} It is a logical possibility that the R-charge of the
$p$'th meson may cross the line $R=2/3$ more then once. Then one
would have to switch back to the description in terms of $\al^{(p-1)}$
when this happens the second time. The results presented in figures 1, 2
indicate that this possibility is not realized, since $R(Q)$ and $R(X)$
are positive and monotonically decreasing functions of $x$.
\item{(3)} The R-charge $R(Q)$ (and consequently $R(X)$) obtained using the
algorithm described above is different from that obtained from  \naivR, $R^{(0)}(Q)$.
For example, the asymptotic value of $R(Q)$ as $x\rightarrow\infty$ is (see Appendix A)
\eqn\RQour{y_{as}={\sqrt{3}-1\over3}\simeq 0.24402~,
}
while \naivR\ leads to
\eqn\RQnaive{y'_{as}={3-\sqrt 5\over 3}\simeq 0.25464~.}

\midinsert\bigskip{\vbox{{\epsfxsize=3in
        \nobreak
    \centerline{\epsfbox{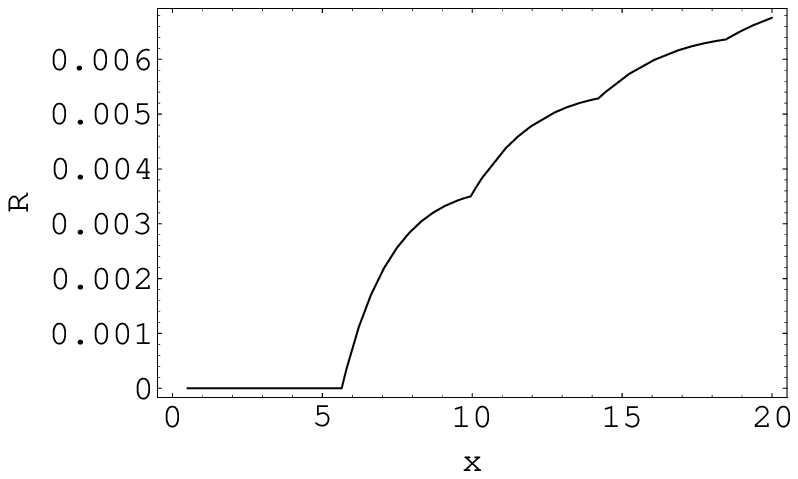}}
        \nobreak\bigskip
    {\raggedright\it \vbox{
{\bf Fig 3.}
{\it  The difference $R_0(Q)-R(Q)$ as a function of $x$. This function is not smooth
at the points where the mesons $\MM_p$ $ (p=1,2,\cdots)$ decouple. }}}}}}
\bigskip\endinsert

\noindent
In figure 3 we plot the difference between $R_0(Q)$ and $R(Q)$. Comparing figures 2 and
3, we see that the quantitative difference between the two is rather small (at most at the
few percent level).

To conclude this section we would like to discuss 
the implications of our analysis for the $a$-theorem.
The latter predicts that for $x>1/2$,
\eqn\alem{{\al}_{UV}>{\al}_{IR}>0~.}
Since the UV theory is free, one has 
\eqn\auv{\al_{UV}={2\over9}N_c N_f\cdot2+{2\over9}N_c^2+2N_c^2=
N_f^2\left[{4\over 9}x+{20\over 9}x^2\right ]~,
}
where we listed the contribution of the fundamentals $Q$, $\tilde Q$, followed by that of the
adjoint field $X$, and of the gauginos.

\midinsert\bigskip{\vbox{{\epsfxsize=3in
        \nobreak
    \centerline{\epsfbox{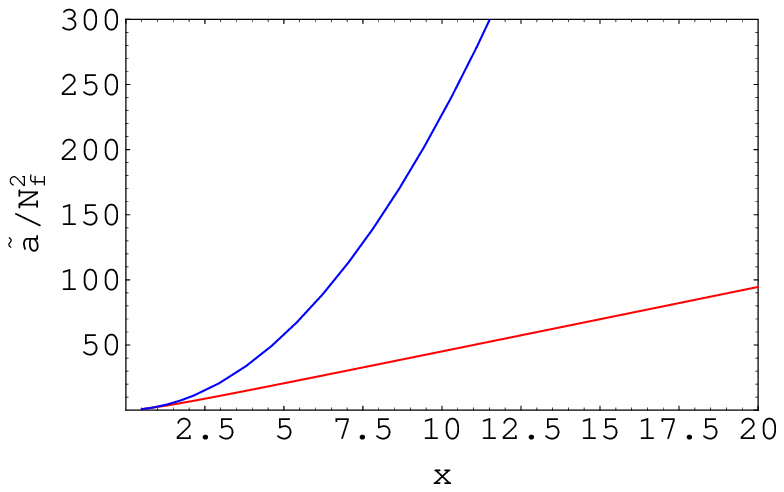}}
        \nobreak\bigskip
    {\raggedright\it \vbox{
{\bf Fig 4.}
{\it The UV (top, blue curve) and IR (bottom, red curve) values of the
central charge  $\al/N_f^2$ as a function of $x$. }}}}}}
\bigskip\endinsert
In figure 4 we plot ${\al}_{UV}$ and ${\al}_{IR}$. We see that \alem\ is indeed satisfied.
Figure 4 also shows that, as seen in the asymptotic large $x$ analysis of
appendix A, $\tilde a_{IR}$  does not contain a term quadratic in $x$ as
$x\to\infty$, in contrast to $\tilde a_{UV}$ \auv. By fitting the data leading
to figure 4 to an asymptotically linear function, one finds
\eqn\asatilde{\tilde a_{IR}/N_f^2=c x+d +O(1/x)~,}
where
\eqn\cdnum{c\simeq 4.976;\;\; d\simeq -5.08~.}
The value of $c$ agrees with the analytic result obtained in
appendix A, $c=4(2+\sqrt{3})/3$. The fact that $d$ is negative will be seen later
to be a necessary condition for the validity of the $a$-theorem.

\midinsert\bigskip{\vbox{{\epsfxsize=3in
        \nobreak
    \centerline{\epsfbox{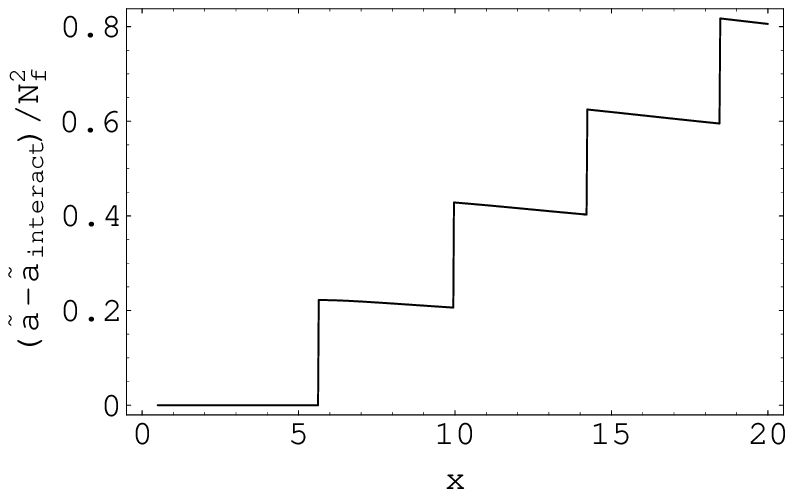}}
        \nobreak\bigskip
    {\raggedright\it \vbox{
{\bf Fig 5.}
{\it The difference between the full central charge, $\al$, and the 
contribution of the interacting part of the IR SCFT, $\al_{interact}$, 
as a function of $x$. Discontinuities correspond to points where 
additional mesons become free.}}}}}}
\bigskip\endinsert

A slightly more sensitive test of the positivity of $\tilde a$ than figure 4
is to compute only the contribution of the interacting part of the IR CFT,
which can be obtained by a generalization of \ainteract
\eqn\atotalgen{\al_{interact}(x)=\al(x)-p\al_M\left({2\over 3}\right)=
\al^{(0)}(x,y(x))-\sum_{j=1}^p \al_M(R(\MM_j))~,}
where $p$ is  the number of mesons which are free at $x$.
We plot the difference of the full central charge and this quantity in figure 5.
Comparing figures 4 and 5 we see that the contribution of 
the interacting part of the CFT is always positive, and in
fact is always much larger than the contribution of the decoupled free fields.

\newsec{Deformations of adjoint SQCD}

In this section we use the improved understanding of the infrared
behavior of adjoint SQCD discussed in section 2, to study
perturbations of that fixed point. There are two different
types of deformations that one can consider: giving v.e.v.'s
to massless scalars with vanishing potentials, and perturbing
the Lagrangian by relevant operators.
We will next consider these two types of perturbations in turn.

\subsec{Higgsing}

Adjoint SQCD has a classical moduli space whose (complex) dimension
is $2N_fN_c$. It is believed that the quantum theory has a moduli space
of the same dimension. Here we will discuss a particular subspace of the
moduli space, corresponding to turning on expectation values of the
adjoint superfield $X$,
\eqn\xvev{\langle X\rangle={\rm diag} (\alpha_1^{n_1}, \alpha_2^{n_2},\cdots,
\alpha_l^{n_l});\;\;\sum_j n_j=N_c~,}
\ie\ the first $n_1$ eigenvalues of $X$ are equal to $\alpha_1$, the
next $n_2$ are equal to $\alpha_2$, etc. By definition,
$\alpha_j$, $j=1,2,\cdots, l$ are all distinct.
The v.e.v. \xvev\ corresponds to a relevant deformation of the theory; it
leads in the infrared to a direct product of $SU(n_j)$ adjoint SQCD
theories\foot{There are also some $U(1)$ factors in the gauge group,
but these can be ignored in the large $N$ limit \limit.},
each of which flows to a fixed point of the sort discussed in section 2.
The corresponding central charge $\tilde a$ is the sum of the central
charges of the individual factors. Defining the variables $x_j=n_j/N_f$,
in analogy to \limit, the $a$-theorem implies that
\eqn\ahiggs{\tilde a(x) > \sum_{j=1}^l\tilde a(x_j)~,}
where $x_j>0$ and
\eqn\sumxj{ \sum_{j=1}^l x_j=x~.}
Equation \ahiggs\ is a non-trivial  requirement on the function $\tilde a$.
It is not clear to us how to prove or disprove it in general;
we next check its validity in a few cases.

Consider, for example, the symmetry breaking pattern $SU(N_c)\to
[SU(N_c/l)]^l$ corresponding to $n_1=n_2=\cdots=n_l=N_c/l$.
In this case all the $x_j$ are equal to each other, and by \sumxj,
$x_j=x/l\equiv b$. The condition
\ahiggs\ is now
\eqn\aahh{\tilde a(x)>{x\over b}\tilde a(b)~.}
Since $x$ can be taken to be arbitrarily large, comparing
to \asatilde\ we see that the $a$-theorem leads to the
constraint
\eqn\atildeb{{\tilde a(b)\over bN_f^2}< c=4(2+\sqrt{3})/3~,}
for all $b$. The simplest case is $b<1/2$. In this case, the $SU(N_c/l)$
theories one finds in the infrared are not asymptotically free,
and their central charge is given by the free field theory result,
\auv:
\eqn\nonfree{\tilde a(b)/N_f^2={4\over9}b+{20\over9}b^2~.}
The constraint \atildeb\ now reads
\eqn\constbfree{{4\over9}+{20\over9}b<4(2+\sqrt{3})/3~.}
It is indeed valid for $b<1/2$, the region of validity of \nonfree.
It would have been violated when $b\simeq 2$, but in that
regime one has to use \aattii\ for $\tilde a(b)$, and one can check
that it too satisfies the constraint \atildeb. 
More generally, one can check  that \atildeb\ is satisfied for all $b$ (see figure 6).

\midinsert\bigskip{\vbox{{\epsfxsize=3in
        \nobreak
    \centerline{\epsfbox{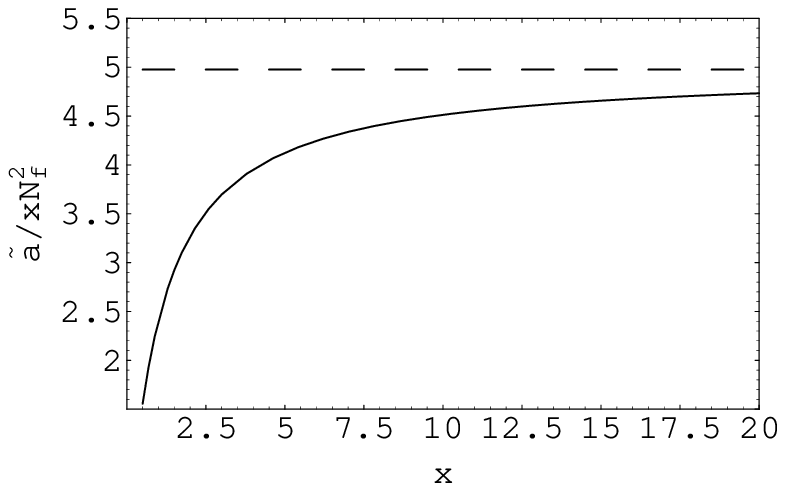}}
        \nobreak\bigskip
    {\raggedright\it \vbox{
{\bf Fig 6.}
{\it $\tilde a(x)/ x N_f^2$ (solid line) is bounded from above by $c=4(2+\sqrt{3})/3$
(dashed line), and approaches it for large $x$. This agrees with the expectation from
the $a$-theorem, \atildeb.}}}}}}
\bigskip\endinsert

Another simple check of \ahiggs\ is obtained by considering
the case where both $x$ and the $x_j$ are very large. In that
case one can use the asymptotic formula \asatilde\ both for the 
left and right hand sides of \ahiggs:
\eqn\checkcd{cx+d>\sum_{j=1}^l(cx_j+d)~.}
The leading term (linear in $x$) is the same on both sides (see \sumxj).
The inequality is indeed satisfied since $d$ is negative (see \cdnum).
A positive $d$ would lead to a violation of the $a$-theorem;
if $d$ vanished, one would have had to consider the $O(1/x)$
term in \asatilde.

\subsec{Relevant superpotential perturbations}

Another interesting class of deformations of adjoint SQCD
corresponds to relevant perturbations of the superpotential
(F terms). These perturbations can be expressed in terms of
the chiral operators in eq. \ginv. The baryons (first line of \ginv) are
in general irrelevant and can be ignored. We will focus here
on perturbations by the operators on the second line of \ginv,
\eqn\wel{
W_k(X)=g_k\tr X^{k+1}~.
}
It would be interesting to generalize the discussion to perturbations
by the mesons on the third line of \ginv, $\MM_j$. 

As discussed in the introduction, although the 
deformations \wel\ are irrelevant in the UV for all
$k>1$ (the case $k=1$
corresponds to a mass term for $X$), they might
lead to relevant deformations of the IR fixed point
of adjoint SQCD. Indeed, we saw in section 2 that
as we vary $x$ \limit, the $R$-charge of $X$ 
monotonically decreases, approaching zero at large
$x$ as
\eqn\asrx{R(X)\simeq {4-\sqrt{3}\over 3x}~.}
Therefore, for any given $k$ there always exists an $x_k$
such that
\eqn\xrel{R(X^{k+1})\leq 2,\qquad {\rm for}\,\,x\geq x_k~.}
For $x>x_k$, $\tr X^{k+1}$ is a relevant deformation
of the infrared fixed point and by turning it on
we expect to flow to a new  fixed point. Assuming that
the R-charge at this fixed point should be visible throughout
the RG flow from the UV, one can determine it using \nfcrq\ 
and the condition that $W_k$ is marginal
\refs{\KutasovVE,\KutasovNP}, 
\eqn\rcharg{
y_k=R_k(Q)=R_k(\tilde Q)=1-{2\over k+1}x,\qquad R_k(X)={2\over k+1}~.
}
It was shown in \KutasovNP\ that the theory with the perturbation
\wel\ turned on has a stable vacuum only for $N_f\ge N_c/k$, or
\eqn\vac{ x\le k~.}
This was done by deforming the superpotential \wel\ to a generic polynomial
of degree $k+1$ in $X$, using the results of \refs{\AffleckMK,\AffleckXZ} 
to show that the resulting  model has no vacuum for
$x>k$, and then removing the perturbation and going back to \wel.

Implicit in the argument of \KutasovNP\ was the assertion that the
perturbation \wel\ must become relevant before one reaches the
point $x=k$, where it destabilizes the theory. Indeed, it would be
inconsistent for a perturbation which is irrelevant at the IR fixed point
of the gauge theory to destabilize that fixed point at long distances.
Therefore, we conclude that it must be that (see \xrel)
\eqn\xxkk{x_k<k~.}
At the time \KutasovNP\ was written, $x_k$ was not known, but now,
using the results of \IntriligatorJJ\ and section 2 of this paper, we can
verify \xxkk. For small $k$, one can use \naivR\ and \xrel, which lead to
\eqn\xk{x_k=\sqrt{{1\over 20}\left({(5k-4)^2\over 9}+1\right)}~.}
For example, for $k=2$, which corresponds to a cubic superpotential
\wel, one finds $x_2=1/2$. Indeed, this perturbation is marginally
irrelevant at the UV fixed point of the gauge theory, but is relevant
at the IR fixed point whenever the gauge theory is asymptotically free.

Recall that \naivR\ is valid only for $x\leq 3+\sqrt 7$. This implies
that \xk\ is only valid for $k\leq 15$. For larger $k$, the full analysis
of section 2 is needed. For  $k>>1$, using the asymptotic form of the
solution, \asrx, one finds
\eqn\xkasymp{x_k\simeq{4-\sqrt{3}\over6}k~.}
It is easy to verify that both \xk\ and \xkasymp\ satisfy the
constraint \xxkk. In fact, since the quantitative difference between
the IW result \xk\ and the exact result is small, one can use \xk\
to verify \xxkk\ for all $k$.

One can actually provide a better bound on $x_k$ as follows.
The fact that the R-charges $R(Q)=y$ and $R(X)=(1-y)/x$ are monotonically
decreasing functions of $x$ can be used to prove the following inequality
\eqn\ineq{{x_{k+1}\over k+2}>{x_k\over k+1}~.}
Indeed
\eqn\xkeq{{x_{k+1}\over k+2}=
{1-y(x_{k+1})\over 2}>{1-y(x_k)\over 2}={x_k\over k+1}~,}
where we  used the fact that $y$ is a monotonically decreasing function of $x$
(see fig. 2). Furthermore, \ineq, \xkeq\ imply that
\eqn\cons{{x_k\over k+1}<{1-y_{as}\over 2}=\half\left(1-{\sqrt{3}-1\over3}\right)~,}
which is a stronger bound than \xxkk\ and will be useful for other purposes below. Note
that the bound \cons\ is saturated in the limit $k\to\infty$ (see \xkasymp).

We would next like to apply the results discussed above to the different
flows associated with the superpotentials \wel. We will focus on two types 
of flows. The first is the flow from the IR fixed point of adjoint SQCD 
with $W=0$ to the fixed point $\bf k$ obtained by turning on $g_k$ \wel.
The second is the flow from $\bf k$ to $\bf k'$, with $k'<k$. This flow
is obtained by studying a superpotential of the form
\eqn\wweell{W(X)=g_k{\rm tr} X^{k+1}+ g_{k'}{\rm tr} X^{k'+1}~.}
We first set $g_{k'}=0$, flow to the infrared fixed point $\bf k$, and then
turn on $g_{k'}$ to further flow to $\bf k'$. As discussed above, $g_k$
corresponds to a relevant perturbation only for $x>x_k$,
so we should restrict consideration to this range (and take into
account the condition for having a stable vacuum \vac, $x\le k$).

One of the main questions we would like to address is the validity
of the $a$-theorem along these flows. To compute the central charge
$a$, we must determine the gauge invariant chiral operators which
might become free as we vary $x$. The chiral ring of the gauge
theory with the superpotential \wel\ is generated by the operators
\refs{\KutasovVE-\KutasovSS}
\eqn\ginvar{
\eqalign{
&{\cal B}^{(n_1,n_2,\cdots,n_k)}=Q^{n_1}_{(1)}\cdots Q^{n_k}_{(k)};
\qquad \sum_{l=1}^kn_i=N_c;\quad n_i\leq N_f~,\cr
&\tr X^2, \tr X^3,\cdots,\tr X^k~,\cr
&({\cal M}_j)_i^{\tilde i}=Q_i X^{j-1} Q^{\tilde i},\quad j=1,\cdots,k~.
}
}
$Q_{(l)}$ are defined by \qx.
As in section 2, the operators $\tr X^l$ do not contribute
to the corrections to $\al$ in the limit \limit\ because their 
contribution is down by a factor of $N_f^2$.
The baryons $\BB$ need to be examined more carefully than in 
section 2, since as one can
see from \rcharg, $R_k(Q)$ becomes negative for
\eqn\xbound{
x>{k+1\over 2}~,
}
so the R-charge of the baryons, which is given as before by
\eqn\rchrgb{
R({\cal B}^{(n_1,n_2,\cdots,n_k)})=\sum_{i=1}^k n_i(i-1)R_k(X)+N_c R_k(Q)~,
}
can potentially violate the unitarity bound.
We show below that for $x\leq k$ the R-charge of  the baryons is positive
(and infinite in the limit \limit) so there should not be any corrections of type
\ansel\ associated with them. Indeed, plugging  \rcharg\ in \rchrgb\ one finds
\eqn\Rcla{
R({\cal B}^{(n_1,n_2,\dots,n_k)})={2\over k+1}\left[N_c\left({k-1\over 2}-x\right)+\sum_{i=1}^k n_i i\right]~.
}
To find a lower bound on $R({\cal B}^{(n_1,n_2,\dots,n_k)})$, we would like to find a lower
bound on  $\sum_{i=1}^k n_i i$. It is easy to see that
\eqn\ineq{
\sum_{i=1}^k n_i i\geq N_f \sum_{i=1}^{[x]}i+(N_c-[x]N_f)([x]+1)i=N_f {[x]([x]+1)\over 2}+(N_c-[x]N_f)([x]+1)~,
}
where $[x]$ is the integer part of $x$. Indeed, $\sum_{i=1}^k n_i i$ is
minimized by the following choice of the $n_i$
\eqn\none{
n_1=\dots=n_{[x]}=N_f;\quad n_{[x]+1}=N_c-[x]N_f; \quad n_i=0,\,\,{\rm for}\,\, i=[x]+2,\dots k~.
}
Eq. \ineq\ leads to the following bound on the R-charge of the baryons:
\eqn\boun{
R({\cal B}^{(n_1,n_2,\dots,n_k)})\geq{N_f\over k+1}\left\{x(k+1)-[x]-(2x^2+[x]^2-2x[x])\right\}~.
}
The expression on the right hand side of \boun\ is a monotonically decreasing function of $x$ for
$x>(k+1)/ 2$. It vanishes\foot{For $x>k$, even separately from the fact that the quantum
theory has no vacuum, the baryons \ginvar\ do not exist, so the issue of their
R-charge does not arise.} at $x=k$. We conclude that the R-charges of baryons  are
large and positive, and thus baryons do not contribute to the correction terms in $\al_k$.
The mesons $\MM_j$ do contribute to these correction terms, as in the discussion of
section 2.  Their R-charges are given by
\eqn\mesW{
R({\cal M}_j)=2y_k+(j-1){1-y_k\over x}=2{j+k-2x\over k+1}~.
}
Plugging into \ansel\ one finds
\eqn\correctk{
\eqalign{
&\al_k(x)/N_f^2=6x(y_k-1)^3-2x(y_k-1)+3x^2\left({1-y_k\over x}-1\right)^3-x^2\left({1-y_k\over x}-1\right)+2x^2+\cr
&\cr
&{1\over 9}\sum_{j=1}^{p(x)}
\left[2- 3\left(2y_k + (j - 1){1 - y_k\over x}\right)\right]^2\left[5 - 3\left(2y_k + (j - 1){1 - y_k\over x}\right)\right]
=\cr
&{4\over (k+1)^3}\left[x^2(2+k+5k^2-12x^2)-{1\over 9}\sum_{j=1}^{p(x)}(-5+6j+k-12x)(1-3j-2k+6x)^2\right]~.}
}
Here $p(x)$ is the number of mesons which are free at $x$,
\eqn\px{
p(x)=\left\{\eqalign{&\left[{1\over 3}(6x-2k+1)\right]\quad {\rm if}\, \left[{1\over 3}(6x-2k+1)\right]\leq k\cr
&k\quad{\rm otherwise}}\right\}~,
}
where $[...]$ is the integer part of the expression in brackets if the expression is positive
and $0$ otherwise. The  $p$-th meson becomes free at
\eqn\xel{
x(p)={1\over 6}{(3p+2k-1)}~.
}
Using eq. \xkasymp\ as an estimate for $x_k$, we see that
at $x\simeq x_k$, some mesons have already decoupled (the
precise number depends on $k$).

We are now ready to consider the RG flows associated with the polynomial superpotentials
\wel, \wweell. Consider first the flow from the infrared fixed point of adjoint SQCD to
the fixed point ${\bf k}$ obtained by switching on \wel. The $a$-theorem expectation is
\eqn\athm{\al(x)>\al_k(x)>0,\qquad {\rm for}\,\,k> x> x_k~.}
As pointed out in \IntriligatorJJ,  \athm\ is guaranteed to be satisfied in a finite region near
$x=x_k$, by construction. Indeed, $\al(x)$ was found in section 2 by maximizing $a^{(p)}(x,y)$
\correct\ with respect to $y$. $\al_k(x)$ is obtained by fixing $y$ to a particular value \rcharg.
The two coincide at $x=x_k$. Therefore, if $x$ is sufficiently close to $x_k$, \athm\ is satisfied.
If $x$ deviates significantly from $x_k$, this argument no longer applies, since the maximum
leading to $\al(x)$ is a {\it local} one.
\midinsert\bigskip{\vbox{{\epsfxsize=3in
        \nobreak
    \centerline{\epsfbox{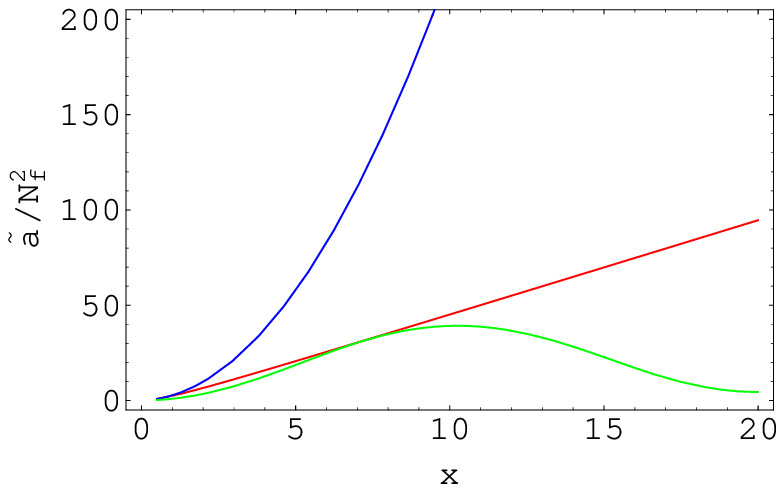}}
        \nobreak\bigskip
    {\raggedright\it \vbox{
{\bf Fig 7.}
{\it The central charge corresponding to the UV (top, blue curve) and IR (middle, red curve)
fixed points of the theory with vanishing superpotential, and that of the fixed point $\bf k$
with $k=20$ (bottom, green curve), as a function of $x$.}}}}}}
\bigskip\endinsert

We have not found a  proof of \athm, but our numerical results suggest that
it is always satisfied. In figure 7 we exhibit the typical behavior of the
central charges corresponding to the free UV fixed  point, $\al_{UV}(x)$,
the IR fixed point of adjoint SQCD with $W=0$, $\al(x)$, and the fixed point
$\bf k$ associated with \wel, $\al_k(x)$, for $k=20$. The UV and IR curves for
the $W=0$ problem are as in section 2 (see figure 4).
The point at which the two lower curves meet is $x=x_k$. As we saw earlier
(eq. \cons), $1/ 2<x_k<(k+1)/2$. The difference $\al(x)-\al_k(x)$ is plotted
in figure 8.
\midinsert\bigskip{\vbox{{\epsfxsize=3in
        \nobreak
    \centerline{\epsfbox{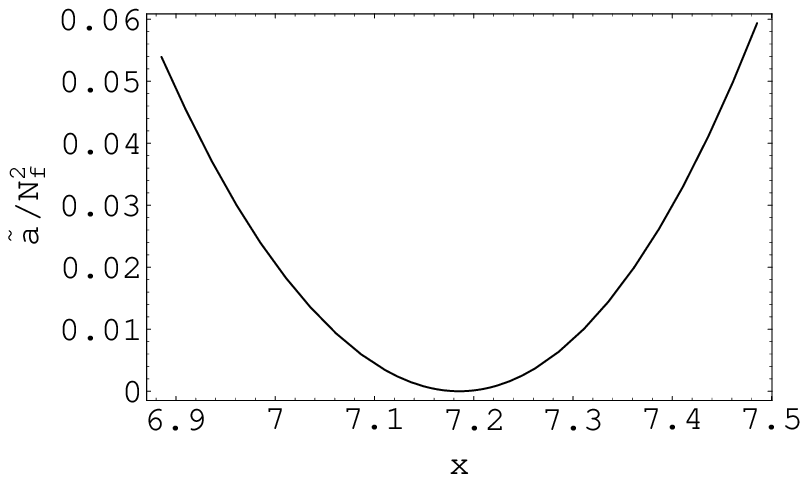}}
        \nobreak\bigskip
    {\raggedright\it \vbox{
{\bf Fig 8.}
{\it The difference $(\al(x)-\al_k(x))/N_f^2$ as a function of $x$; $k=20$. }}}}}}
\bigskip\endinsert

Note that in figures, 7,8 we have extended the curve $\al_k(x)$ to the full range
$1/2<x<k$. This is of course unphysical, since as discussed earlier, the fixed point
$\bf k$ does not exist for $x<x_k$. Thus, the lower curve in figure 7 is unphysical
to the left of the point where it touches the curve above it. Similarly, the part of the
curve of figure 8 to the left of the point where it touches the x axis is unphysical.
At any rate, the main conclusion from figures 7,8 is that the $a$-theorem seems
to be satisfied for these flows.

\midinsert\bigskip{\vbox{{\epsfxsize=3in
        \nobreak
    \centerline{\epsfbox{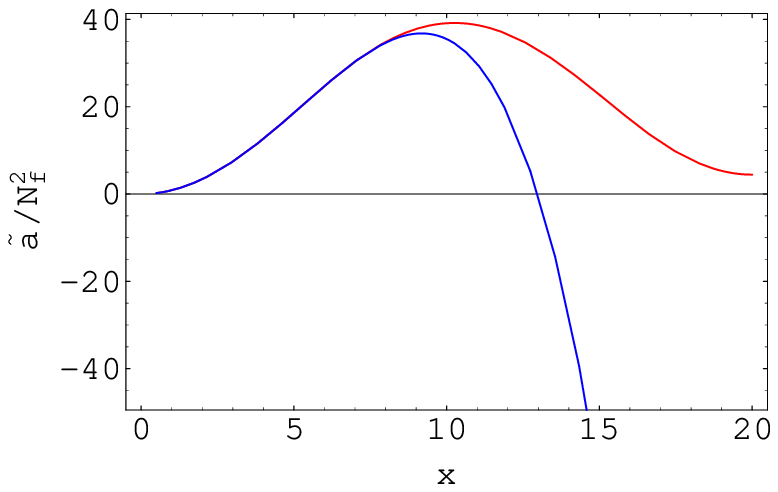}}
        \nobreak\bigskip
    {\raggedright\it \vbox{
{\bf Fig 9.}
{\it The central charge corresponding to the fixed point $\bf 20$ associated
with the superpotential ${\rm tr} X^{21}$. 
The top (red) curve takes into account
the corrections due to the decoupling of meson fields discussed in the text,
while the bottom (blue) curve neglects these corrections.}}}}}}
\bigskip\endinsert
It is worth noting that the corrections \ansel\ are crucial for the positivity of $\al_k$.
In figure 9 we plot the uncorrected central charge, $\al^{(0)}_k$, together with the
corrected one, $\al_k$. The former is actually  negative in part of the physical domain
$x_k<x<k$, while the latter is positive there.

We next turn to the flows ${\bf k}\to{\bf k'}$ associated with superpotentials
of the form \wweell. The $a$-theorem predicts that
\eqn\afk{\al_k>\al_{k'},\qquad{\rm for}\;\; x>x_k~.}
The authors of \AnselmiYS\ pointed out that in general,
this prediction seems to be violated in a range of $x$'s:
\eqn\athviol{\al_k<\al_{k'},\qquad{\rm for}\;\;\half<x<x_{\rm int}(k,k')~.}
They computed the value of $x_{\rm int}(k,k')$ in a few examples, obtaining
the results
\eqn\xintsmall{\eqalign{
x_{\rm int}(3,2)\simeq & 0.65~,\cr
x_{\rm int}(4,3)\simeq & 1.01~,\cr
x_{\rm int}(5,4)\simeq & 1.38~.\cr
}}
Thus, the $a$-theorem can potentially be violated in these flows, depending
on the value of $x_k$. Comparing equations \afk\ and \athviol, we see that a necessary
condition for the validity of the $a$-theorem is
\eqn\condxkk{x_k>x_{\rm int}(k,k')~,}
for all $k$, $k'<k$.
At the time \AnselmiYS\ was written, $x_k$ was not known and thus
it was impossible to check \condxkk. Now, we can use the results of
\IntriligatorJJ\ and this paper to do that. For the cases \xintsmall,
which involve small values of $k$, one can use \xk:
\eqn\xthfrfi{\eqalign{
x_3\simeq & 0.85~,\cr
x_4\simeq & 1.21~,\cr
x_5\simeq & 1.58~.\cr
}}
Comparing to \xintsmall\ we see that the necessary condition \condxkk\
is satisfied in all three cases.

Our numerical results seem to suggest that \condxkk\ is always satisfied.
As an illustration, in fig. 10 we plot the behavior of the various central
charges for the case $k=8$, $k'=3$. The top curve is the central charge
of the IR fixed point of the theory with vanishing superpotential. 
It intersects the two curves corresponding to the fixed points associated 
with the superpotentials ${\rm tr} X^4$ and ${\rm tr} X^9$ at the points 
$x=x_3$ and $x=x_8$, respectively (recall that $x_8>x_3$; this can be 
used to determine which
curve is which in fig. 10). We see that $a_8$ is larger than $a_3$, in
agreement with the $a$-theorem, for all $x>x_8$. The point at which
the two curves cross, $x_{\rm int}(8,3)$ lies between $x_3$ and $x_8$
and does not lead to conflict with the $a$-theorem, since the ${\rm tr} X^9$
superpotential is irrelevant for $x<x_{\rm int}(8,3)$, and so the
RG flow in question does not exist in that regime.

\midinsert\bigskip{\vbox{{\epsfxsize=3in
        \nobreak
    \centerline{\epsfbox{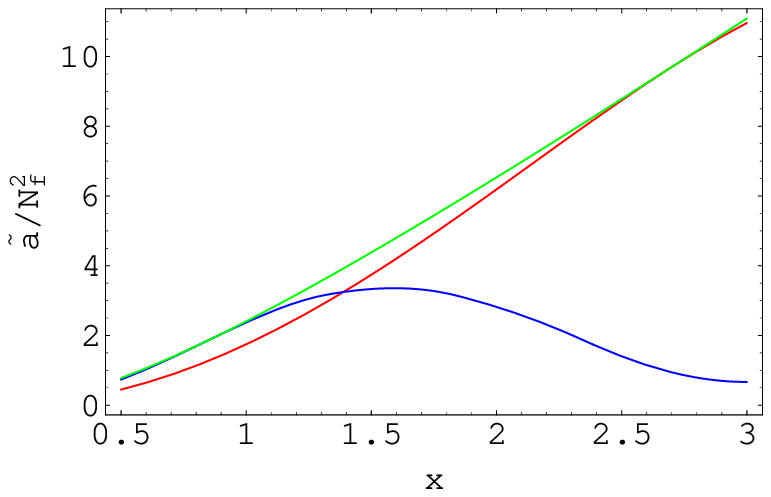}}
        \nobreak\bigskip
    {\raggedright\it \vbox{
{\bf Fig 10.}
{\it The central charges associated with the infrared fixed point of adjoint
SQCD with vanishing superpotential (green), the fixed point $\bf 8$ associated
with superpotential ${\rm tr} X^9$ (red), and the fixed point $\bf 3$ 
associated with ${\rm tr} X^4$ (blue). }}}}}}
\bigskip\endinsert

All other examples of pairs  $k$, $k'$ that we checked, behave
in a qualitatively similar way.

\newsec{Strong-weak coupling duality}

In sections 2 and 3 we used the assumption that the $U(1)_R$ current that
becomes part of the superconformal algebra at non-trivial fixed points of the
RG is visible as an anomaly free symmetry throughout the RG flow. As we
discussed in the introduction, this assumption may fail at strong coupling.
There is no known a priori way to determine when that will happen, but
one way to discover that it does is to use Seiberg duality.

It is thus natural to wonder whether the results of sections 2 and 3 should be
modified at strong coupling. For the case of adjoint SQCD with
vanishing superpotential, discussed in section 2, there is no known Seiberg-type 
duality, and so no tools for addressing this question at present. As we saw in 
sections 2,3, and will see further in this section, one gets a consistent picture 
by assuming that no modifications of the results in section 2 are necessary, but 
that of course does not imply that they are correct for all $N_f$. It cannot be
excluded that for $N_f<N_f^*$ (with some $N_f^*>0$) the formulae of section 2 are 
no longer valid, but we have not found any evidence for this in our work.

In this section, we will discuss the fixed point $\bf k$ obtained by perturbing adjoint
SQCD by the superpotential \wel\ (for $x>x_k$ \xrel), where a dual description
is known to exist \refs{\KutasovVE-\KutasovSS}, and one can ask what it
predicts for the properties of the fixed point $\bf k$ at strong coupling.

The duality of \refs{\KutasovVE-\KutasovSS} relates adjoint SQCD with
gauge group $SU(N_c)$ and superpotential \wel\  to an $\NN=1$ supersymmetric
gauge theory with gauge group $SU({\tilde N}_c)=SU(kN_f-N_c)$ and the following 
matter content: an adjoint field $Y$, $N_f$ chiral superfields $q_i$, 
$\tilde q^{\tilde i}$ in the anti-fundamental and fundamental representation of the
gauge group, respectively, and gauge singlets $(M_j)^i_{\tilde i}$, $j=1,\cdots,k-1$.
The  superpotential of this  theory is given by
\eqn\wmag{
W_{mag}=-g'_k\tr Y^{k+1}+{1\over \mu^2}g'_k\sum_{j=1}^{k}M_j\tilde q Y^{k-j}q~,
}
where $\mu$ is an auxiliary scale.
We will refer to these two  theories as electric and magnetic, respectively.
The conjecture of \refs{\KutasovVE-\KutasovSS} is that they flow in the infrared
to the same fixed point. The operator matching between the electric and magnetic theories
is\foot{We neglect numerical proportionality constants.}:
\eqn\match{\eqalign{&({\MM_j})_{\tilde i}^i=
{\tilde Q}_{\tilde i}X^{j-1}Q^i\longleftrightarrow (M_j)^i_{\tilde i}~,\cr
&\Tr X^j\longleftrightarrow \Tr Y^j~,\cr
&\BB_{el}^{(n_1,n_2,\cdot,n_k)}\longleftrightarrow \BB_{mag}^{(m_1,m_2,\cdot,m_k)};\,\,
m_l=N_f-n_{k+1-l};\,\, l=1,2,\cdots k~,
}}
where the magnetic baryons are defined in the same way as the electric ones \ginvar\
(with the substitution $N_c\rightarrow {\tilde N}_c =kN_f-N_c$).

In the next subsection we briefly discuss the magnetic theory
from the point of view of the analysis of sections 2, 3. In subsection
4.2 we discuss the implications of its properties for the electric theory.

\subsec{RG flows in the magnetic theory}

It will be convenient to introduce the magnetic dual of $x$,
\eqn\xequiv{
  \x \equiv { {\tilde N}_c \over N_f}=k-x~.
}
We will mainly discuss the region
\eqn\xdphys{
x, \x \in \left ({1 \over 2},k-{1 \over 2}\right)~,
}
in which both the electric and the magnetic theories are asymptotically free.

When $\x$ is close to (and above) $1/2$, most of the terms in the
superpotential \wmag\ are irrelevant. The only exception
is the term $M_k\tilde q q$, which is relevant in the infrared
fixed point of the magnetic adjoint SQCD for all $\x>1/2$.
This term drives the theory to a new fixed point, at which
$M_1,\cdots, M_{k-1}$ are free but $M_k$ is interacting.
The R-charges and central charge $\tilde a^{m}$ at this fixed
point can be determined in a similar way to that employed in
section 2. Denoting the R-charge of $q,\tilde q$ by $\tilde y$, one has
$R(Y)=(1-\tilde y)/\x$ and $R(M_k)=2-2\tilde y$. Plugging
these charges into the expression for $\tilde a$, and maximizing
w.r.t. $\tilde y$, one can determine $\tilde y$ and $\tilde a^{m}$.

When $\x$ increases further, more and more of the terms in the
superpotential \wmag\ become relevant and have to be taken into
account. To solve for the R-charges, one has to discuss separately
two different ranges of $\x$:
\item{(1)} The ${\rm tr} Y^{k+1}$ term in \wmag\ is irrelevant in the
infrared fixed point of the magnetic adjoint SQCD. The last $p$ meson
fields, $M_j$ with $j=k-p+1,\cdots, k$ are interacting, while the rest of
the meson fields are free. In this case, the R-charge of the interacting mesons
is given by
\eqn\rmj{
R(M_j)=2-2 R(q)-(k-j) R(Y)=2-2\y-(k-j){1-\y \over \x}~.
}
$\y$ is determined by computing the magnetic central charge
as a function of $\y$ and maximizing it.
\item{(2)} The ${\rm tr} Y^{k+1}$ term in the magnetic superpotential
is relevant. In this case we set $R_k(Y)=2/(k+1)$, $y_k=1-2\x/(k+1)$,
and determine the R-charges of the mesons by using \rmj. If \rmj\ gives an
R-charge smaller than $2/3$ to a meson, the corresponding term in the magnetic
superpotential is irrelevant in the infrared, and this meson remains free there.

\noindent
In practice, one proceeds in a way similar to that employed in sections 2 and 3.
Introduce an auxiliary quantity $\al^{m,(p)}$, the magnetic central charge
computed with the assumption that the last $p=(0,1,2,...)$ meson fields
$M_k,\cdots,M_{k-p+1}$ {\it are not} free; their R-charges are given by \rmj.
\eqn\amfna{\eqalign{
&\al^{m,(p)}/N_f^2
=6\x(\y-1)^3-2\x(\y-1)+3\x^2\left({1-\y\over \x}-1\right)^3-
\x^2\left({1-\y\over \x}-1\right)+2\x^2+\cr
&\cr
&{1\over 9}\sum_{j=1}^p
\left[2- 3\left(2\y + (j - 1){1 - \y\over \x}\right)\right]^2
\left[5 - 3\left(2\y + (j - 1){1 - \y\over \x}\right)\right]+{2\over 9}(k-2p)~.
}}
Start with $\al^{m,(1)}$and maximize it w.r.t. $\y$. Denote the value of
$\y$ at the maximum by  $\y^{(1)}(x)$. Vary $\x$ to the point where the $R$-charge
of $M_{k-1}$ \rmj\ approaches $2/3$. At that point the term $M_{k-1}\tilde qYq$
in the magnetic superpotential \wmag\ becomes relevant, and one should switch to
the $\al^{m,(2)}$ description. This  can be continued to arbitrarily large
$\x$.

The above procedure is valid as long as the polynomial superpotential for $Y$ is irrelevant
in the infrared. Define $\tilde x_k$ via the condition
\eqn\xrelm{R(Y^{k+1})=2,\qquad{\rm at}\,\,\x=\x_k~.}
For $\x>\x_k$, one should set the R-charges to the values given in
point (2) above.
\midinsert\bigskip{\vbox{{\epsfxsize=3in
        \nobreak
    \centerline{\epsfbox{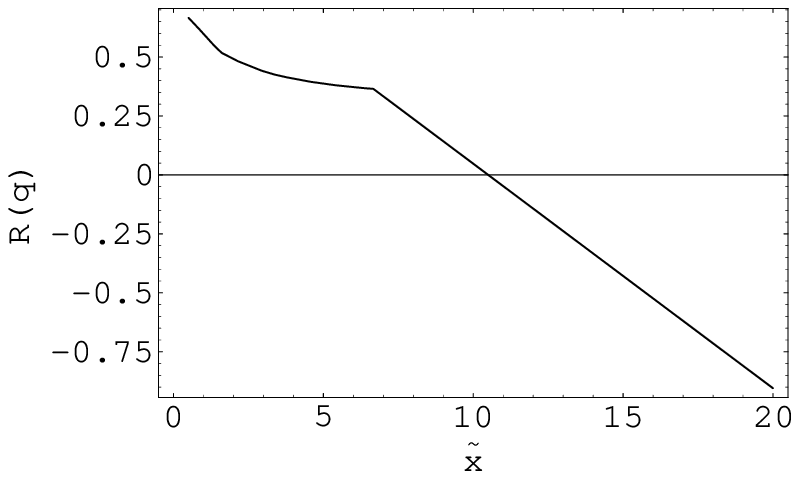}}
        \nobreak\bigskip
    {\raggedright\it \vbox{
{\bf Fig 11.}
{\it R-charge of $q$ as a function of $\x$ at $k=20$. }}}}}}
\bigskip\endinsert
\midinsert\bigskip{\vbox{{\epsfxsize=3in
        \nobreak
    \centerline{\epsfbox{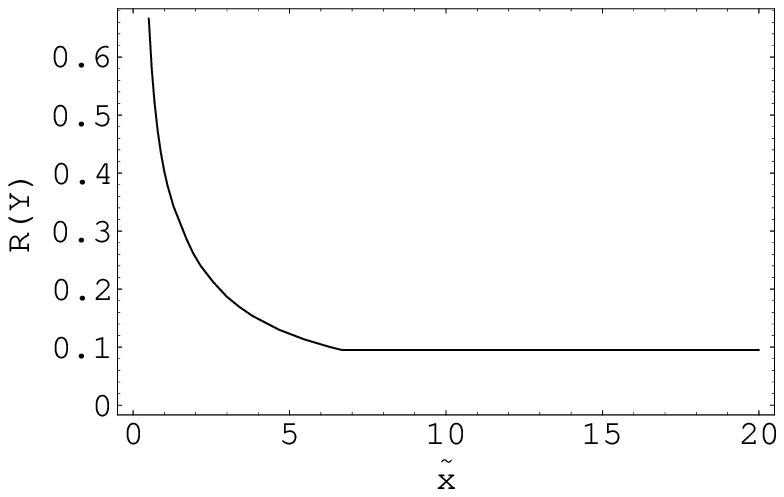}}
        \nobreak\bigskip
    {\raggedright\it \vbox{
{\bf Fig 12.}
{\it R-charge of $Y$ as a function of $\x$ at $k=20$.  }}}}}}
\bigskip\endinsert
\midinsert\bigskip{\vbox{{\epsfxsize=3in
        \nobreak
    \centerline{\epsfbox{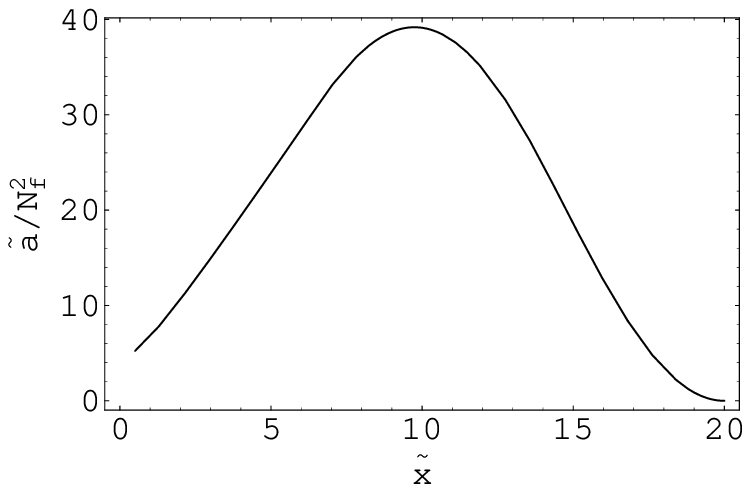}}
        \nobreak\bigskip
    {\raggedright\it \vbox{
{\bf Fig 13.}
{\it $\am_k$ as a function of $\x$ at $k=20$. }}}}}}
\bigskip\endinsert
\midinsert\bigskip{\vbox{{\epsfxsize=3in
        \nobreak
    \centerline{\epsfbox{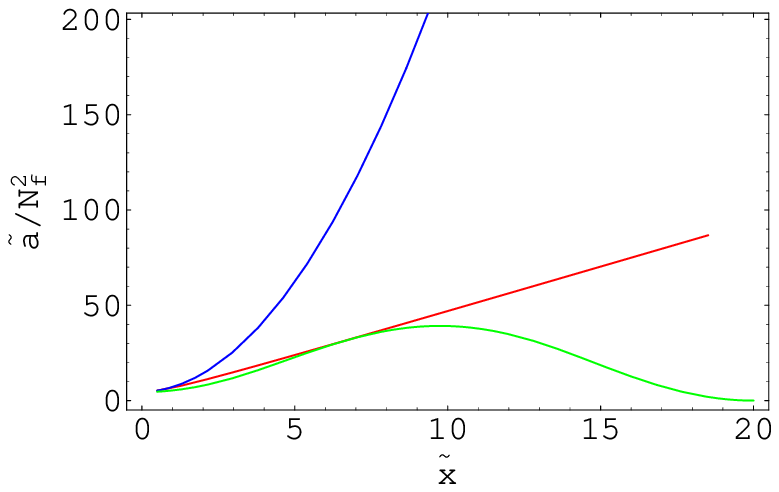}}
        \nobreak\bigskip
    {\raggedright\it \vbox{
{\bf Fig 14.}
{\it The central charge as a function of $\x$ corresponding to the UV (top, blue curve)
and IR (middle, red curve) fixed points of the magnetic theory with the coupling in front
of $\Tr Y^{k+1}$ tuned to zero, and that of the fixed point with the superpotential \wmag\
turned on (bottom, green curve),  for $k=20$. }}}}}}
\bigskip\endinsert
A few comments about the above procedure:
\item{(a)} The expression \amfna\ looks very similar to the analogous
expression in the electric theory \correct. In particular, the value of $\y$
at the maximum for a given $p$, $\y^{(p)}(x)$ can be obtained from the
one found in section 2 by replacing $x\rightarrow \x$ and $y\rightarrow\y$.
Nevertheless,  the magnetic central charge $\am$ is not related to the electric
one in the same way. The reason is that in the electric theory we saw in section
2 that one switches from the $\tilde a^{(p)}$ to the $\tilde a^{(p+1)}$ description when
\eqn\swite{2y+(p-1){1-y\over x}={2\over 3}~,}
while in the magnetic theory the analogous condition is
\eqn\switm{2\y +(p-1){1-\y\over\x}={4\over 3}~.}
\item{(b)} One may wonder whether \amfna\ should be further corrected
to take into account the decoupling of some other chiral superfields. It is
easy to see that the answer is no. The operators ${\rm tr}Y^j$ can be neglected
for the same reasons as in the electric theory (their contribution to $\tilde a^{(m)}$
is down by a factor of $N_f^2$ from the leading terms). The baryons have a large
positive R-charge, as in the discussion of sections 2,3. Finally, the operators
$\tilde q Y^{j-1} q$ are not chiral, due to the superpotential \wmag.
\item{(c)} In the same way as in section 3, one can derive a bound analogous
to \cons:
\eqn\ccoonnmm{{\x_k\over k+1}<{1-\y_{as}\over2}~,}
where $\y_{as}$ is computed in appendix A.

\noindent
The results of our numerical implementation of the above procedure are depicted in figures 11-14.
Figures 11 and 12 are the plots of the R-charges of $q$ and $Y$  as a function of $\x$ at $k=20$.
For $\x<\x_k$ these R-charges are determined via the maximization procedure, while for $\x>\x_k$
they are determined by the superpotential \wmag. Figure 13 exhibits the magnetic central charge
$\am$ as a function of $\x$. Finally figure
14 is the analog of figure 7 for the magnetic case. It shows the UV central charge, $\am_{UV}$,
which is given by
\eqn\amuv{\am_{UV}=N_f^2\left[{4\over 9}\x+{20\over 9}\x^2+{2k\over 9}\right]~,}
the infrared central charge of the theory corresponding to the superpotential \wmag\
with the  $\Tr Y^{k+1}$ turned off, and the central charge $\am_k$ corresponding to the full
superpotential with $k=20$. As in the electric case, the bottom curve in figure 14 should
only be taken seriously to the right of the point $\x_k$, where it touches the curve above it.

\subsec{Consequences for duality}

One  consequence of the discussion above for the duality of
\refs{\KutasovVE-\KutasovSS} is that for all $k$, there exists an analog
of the ``conformal window'' of supersymmetric QCD \SeibergPQ, where
both the electric and magnetic theories are asymptotically free. In our case
the analogous statement is that there exists a region in $x$, in which  both
the electric  and  the magnetic polynomial superpotentials ${\rm tr} X^{k+1}$
and ${\rm tr} Y^{k+1}$ are relevant in the infrared. Indeed, in section 3 we saw
that ${\rm tr} X^{k+1}$ is relevant for $x>x_k$ where $x_k<(k+1)/2$ (see
eq. \cons). Similarly, the magnetic superpotential is relevant for
$\x>\x_k$ where $\x_k<(k+1)/2$ (see eq. \ccoonnmm). Taking into
account \xequiv, we see that there is a window,
\eqn\confwind{x_k<x<k-\x_k~,}
in which both polynomial superpotentials are relevant.

If $x_k$ and $\x_k$ instead satisfied the inequality
$x_k+\x_k>k$, so that the region \confwind\ did not exist,
the situation would have been much more puzzling.
For $k-\x_k<x<x_k$, the duality of \refs{\KutasovVE-\KutasovSS}
would have then predicted an equivalence of theories in which the
polynomial superpotentials are turned off on both sides. Such a duality
would have had a number of puzzling features, and we view the existence
of the ``conformal window'' \confwind\ as a consistency check on the whole
picture.

The duality of \refs{\KutasovVE-\KutasovSS} implies in particular that the
electric and magnetic R-charges and central charges agree for all $x$,
\eqn\ea{\eqalign{
R_k(X;x)=&\tilde R_k(Y;k-x)~;\cr
\ale_k(x)=&\am_k(k-x)~.}}
It is natural to ask whether equation \ea\ actually agrees with the computations
performed in sections 2 -- 4. The answer is that the two calculations agree in
the conformal window \confwind, but disagree outside of it. The agreement
in the conformal window essentially follows from the anomaly matching
that was checked to hold in the original papers \refs{\KutasovVE-\KutasovSS}.

The disagreement outside the conformal window can be understood even
without detailed calculations. It is clear that the flavor of the calculation is completely
different on the two sides. Consider, for example the region $x<x_k$. The
electric superpotential \wel\ is irrelevant and can be neglected in this region,
and thus the R-charges and central charges should be computed as in section 2,
by maximizing the trial R-charge \naiv. In the magnetic theory, the polynomial
superpotential is strongly relevant in this regime, and thus naively one would expect
that no maximization is necessary, and one just uses $\tilde R_k(Y)=2/(k+1)$, etc.
It would be very surprising if the maximization process of section 2 gave such
a simple, $x$ independent result, and as we saw in section 2, it in fact does not.

A natural interpretation of this disagreement is that, just like in supersymmetric
QCD, for $x<x_k$ the magnetic theory is so strongly coupled that the infrared
R-charge cannot be identified with any symmetry of the UV theory, and the only
way to find the correct answer is to pass to the dual, electric variables. Similarly,
the electric description breaks down for $x>k-\x_k$ and in that region, one has to
use the magnetic variables in order to compute $\al_k$. One can check that doing 
that leads to results consistent with the $a$-theorem.

\newsec{Discussion}

The main motivation for this paper was the recent work of
Intriligator and Wecht \IntriligatorJJ, who proposed a
way to determine the $U(1)_R$ symmetry which belongs to the
$\NN=1$ superconformal algebra at an IR fixed point of a
SUSY gauge theory. Our purpose was to explore in a specific
class of models the interplay between three circles of ideas:
\item{(1)} The $a$-theorem: the conjecture that the combination
of `t Hooft anomalies \adef\ is always lower at an IR fixed
point of the renormalization group than at the corresponding
UV fixed point \refs{\CardyCW - \CappelliDV}.
\item{(2)} Seiberg duality: the conjecture that $\NN=1$ supersymmetric
gauge theories often have the property that two different theories
flow in the infrared to the same fixed point \SeibergPQ. In the models
discussed here, the relevant version of this duality was proposed in
\refs{\KutasovVE - \KutasovSS}. 
\item{(3)} The results of \IntriligatorJJ\ on determining
the R-charge at an infrared fixed point of an $\NN=1$ gauge theory.

\noindent
We showed that the results of \IntriligatorJJ\ (slightly
corrected to take into account unitarity constraints) lead,
in the class of models that we studied, to a more detailed
understanding of the phase structure. The resulting phase
diagram provides some rather non-trivial checks of the
$a$-theorem. 

As discussed in \IntriligatorJJ, the $a$-theorem
is guaranteed to hold when using their results, if the UV and
IR fixed points are sufficiently close to each other. The checks
performed here are non-trivial since many of them are performed
in the opposite regime, where the UV and IR fixed points are very
far from each other. In particular, we showed that the results of
\IntriligatorJJ\ and this paper resolve certain potential problems 
with the $a$-theorem raised in \AnselmiYS. We view the consistency 
of our results with the $a$-theorem as evidence for the 
validity of both.

We also showed that the results of \IntriligatorJJ\ satisfy some
non-trivial consistency conditions with the strong-weak coupling
duality of \refs{\KutasovVE - \KutasovSS}. For example, the fact
that there exists a region in $(N_f,N_c)$ where both the electric
superpotential ${\rm tr} X^{k+1}$, and the magnetic one
${\rm tr} Y^{k+1}$ correspond to relevant perturbations 
of the IR fixed points of the corresponding gauge theories
was necessary for the consistency of \refs{\KutasovVE - \KutasovSS};
the results of \IntriligatorJJ\ and this paper show that
such a region indeed exists. This, too, validates both circles
of ideas. 

The construction of \IntriligatorJJ\ is useful when one can identify
the $U(1)_R$ symmetry of the IR $\NN=1$ superconformal field theory
among the anomaly free symmetries of the full theory. We showed that
this is usually the case in a finite region in parameter space (\eg\
here the space labeled by $x$ \limit), and gave examples in which
this idea fails. In the examples of this failure discussed in this
paper, one can find the right $U(1)_R$ symmetry by switching to a
weakly coupled (Seiberg) dual. It is tempting to conjecture that this
is a general feature.

It might be interesting to repeat the analysis done here in other $\NN=1$
supersymmetric gauge theories, to see whether any difficulties with the
$a$-theorem and/or Seiberg duality arise. The main open problem related
to the subject of this paper is to prove the $a$-theorem.

\bigskip

\noindent
{\bf Acknowledgments:}

We thank K. Intriligator and A. Schwimmer for discussions.
This work was supported in part by DOE grant \#DE-FG02-90ER40560.

\appendix{A}{Large $x$ behavior of the R-charges and the central charge $\tilde a$}

We start by presenting an analytic solution for $\tilde a(x)$ and the R-charges
$R(Q)$ and $R(X)$, for the system studied in section 2, in the large $x$ limit defined
in \limit, \ie\ for $N_f<< N_c$, with both $N_f,N_c\to\infty$.
To do that, we need to implement the procedure described after eq. \correct.
Since we expect the R-charge of $Q$, $y$, to go to a finite constant as $x\to\infty$
(see \eg\ figure 2), it is easy to take the $x\to\infty$ limit of \correct. The first line,
which is nothing but $\tilde a_0(x,y)$ \naiv, is given in the limit by
\eqn\largexao{\tilde a_0(x,y)\simeq N_f^2\left[6x(y-1)^3-10x(y-1)\right]~.}
Note, in particular, that the terms that go like $x^2$ in \naiv\ cancel, and the
leading behavior is linear in $x$.

The second line of \correct\ contains a sum over a large number of terms, since
$p$ (or more precisely $p_0$) diverges in the limit $x\to\infty$. Indeed, to
estimate $p$, one can plug into \mes, and require that $R(\MM_p)=2/3$. The
mistake one makes by doing that is subleading in the $1/x$ expansion. This leads to
\eqn\defplargex{2y+(p-1){1-y\over x}={2\over3}~.}
We see that at large $x$, $p$ is expected to be proportional to $x$.
To perform the sum on the second line of \correct, one can replace $j$
by the variable $t$ defined by
\eqn\deft{t=2y+(j-1){1-y\over x}~.}
At large $x$, $t$ becomes a continuous variable, and one can replace the sum
over $j$ by an integral over $t$. Performing this integral leads to:
\eqn\sumlargex{{1\over 9}\sum_{j=1}^p
\left[2- 3\left(2y + (j - 1){1 - y\over x}\right)\right]^2
\left[5 - 3\left(2y + (j - 1){1 - y\over x}\right)\right]\simeq {N_f^2x\over18}(2-6y)^3~.}
Adding this to \largexao, one finds
\eqn\totala{\tilde a(x,y)\simeq N_f^2x\left[6(y-1)^3-10(y-1)+{1\over18}(2-6y)^3\right]~.}
The local maximum of \totala\ w.r.t. $y$ occurs at
\eqn\ymax{y={\sqrt{3}-1\over3}~,}
in agreement with the numerical results that lead to figure 2. The value of
the central charge $\tilde a$ in this limit is
\eqn\avalue{\tilde a={4N_f^2 x\over3}(2+\sqrt{3})~.}

\noindent
We next move on to the magnetic theory  described in section 4.
To find an analytic expression for $\am$, $R(q)$, $R(Y)$, as $\x\rightarrow\infty$, $\x<\x_k$,
we should implement the procedure outlined after \amfna.
Again as in the case above we expect that the R-charge of $q$, $\y$, goes to a finite
constant as $\x\rightarrow\infty$, so it is easy to take a large $\x$ limit of \amfna.
The first line of \amfna\ behaves exactly in the same way as the
first line in \correct\ and hence is given by \largexao\ (with the replacement $x\rightarrow\x$ and
$y\rightarrow\y$) in the large $\x$ limit.

The second line of \amfna\ contains a sum over a large number of terms,
and a term linear in $p$:  ${2N_f^2\over 9}(k-2p)$.
We can use \switm\ to estimate the value of $p$
\eqn\deflargexxp{p=\left({4\over 3}-2\y\right){\x\over 1-\y}+{\cal O}(1)~.}
To perform the sum we introduce the variable $\tilde t$ defined by
\eqn\defmm{\tilde t=2\y+(j-1){1-\y\over x}~.}
At large $\x$, $\tilde t$ becomes a continuos variable and one can replace the sum over $j$
by an integral over $\tilde t$. The main difference with the case discussed above is that
the limits of integration are different.  While the integral over $t$ is performed over the interval
$(2y,{2\over 3})$, the $\tilde t$ integral runs over the interval $(2y,{4\over 3})$.
Performing the integral we get
\eqn\sumlargex{\eqalign{
&{1\over 9}\sum_{j=1}^p
\left[2- 3\left(2\y + (j - 1){1 - \y\over \x}\right)\right]^2
\left[5 - 3\left(2\y + (j - 1){1 - \y\over \x}\right)\right]\simeq\cr
&{N_f^2\x\over18}(2-6\y)^3+{4N_f^2\x\over 27(1-\y)}~.}}
Adding all the pieces  we obtain the following expression for $\am$
\eqn\totalam{\am=N_f^2\x\left[6(\y-1)^3-10(\y-1)+{1\over 18}(2-6\y)^3+{4(2\y-1)\over 9(1-\y)}\right]
+{2k\over 9}N_f^2~.}
The local minimum of this expression with respect to $\y$ occurs at
\eqn\ymaxm{\y_{as}\simeq 0.2844~.}
The value of central charge in this limit is
\eqn\avaluem{\am\simeq 4.6909N_f^2\x+{2\over 9}kN_f^2~.}
Finally the point $\x_k$, where the coupling \wmag\ becomes relevant is
\eqn\xkm{\x_k\simeq 0.3578k~.}

\listrefs

\bye